\documentclass[preprintnumbers,
amsmath,amssymb,floatfix,10pt,aps,prd,twocolumn,notitlepage,
superscriptaddress,longbibliography,nofootinbib,showkeys]{revtex4-2}
\usepackage{bm}
\usepackage{amsfonts}
\usepackage{latexsym}
\usepackage{amsmath}
\usepackage{palatino}
\usepackage{mathpazo}
\usepackage{textcomp}
\usepackage{float}
\usepackage{booktabs}
\usepackage{dcolumn}
\usepackage{ragged2e}
\usepackage{epsfig}
\usepackage{dsfont}
\usepackage[dvipsnames]{xcolor}
\usepackage{hyperref}
\hypersetup{           
    colorlinks=true,                
    breaklinks=true,                
    urlcolor= red,                
    linkcolor= blue,                
    bookmarksopen=false,
    filecolor=black,
    citecolor= RubineRed,
    linkbordercolor=blue}
\usepackage{graphicx}
\usepackage{orcidlink}
\usepackage[T1]{fontenc}

\begin{document}

\title{Dark Matter Surrounded Quartic Square-root Horndeski Black Hole: Thermodynamics, Optical Properties and Quasinormal Oscillations}
\author{Mrinnoy M. Gohain\orcidlink{0000-0002-1097-2124}}
\email{mrinmoygohain19@gmail.com}
\affiliation{%
 Department of Physics, Dibrugarh University, Dibrugarh \\
 Assam, India, 786004}
 
\author{Kalyan Bhuyan\orcidlink{0000-0002-8896-7691}}%
 \email{kalyanbhuyan@dibru.ac.in}
\affiliation{%
 Department of Physics, Dibrugarh University, Dibrugarh \\
 Assam, India, 786004}%
 \affiliation{Theoretical Physics Divison, Centre for Atmospheric Studies, Dibrugarh University, Dibrugarh, Assam, India 786004}

\begin{abstract}
In this work, we study a special form of Horndeski solutions, viz.  quartic "square-root" Horndeski black hole immersed in a perfect-fluid dark-matter halo, by examining its thermodynamics, null-geodesic shape, optical shadow, and quasinormal ringdown spectrum. The model is characterized by three parameters, namely $\beta$, $\eta$ (non-minimal Horndeski coupling parameters), and $b$ (perfect fluid dark matter parameter), which collectively determine horizon properties and observational effects. To study thermodynamic stability, we used the specific-heat and free-energy arguments, with which we demonstrated that small-horizon states are locally stable but are never globally preferred.  Analytic solutions of null geodesics reveal the radius of the photon sphere and the critical impact parameter, proving that increases in the dark matter parameters and the Horndeski parameter $\beta$ enlarge both the photon sphere and the subsequent shadow, whereas increase in the Horndeski coupling $\eta$ causes a mild diminishing in the shadow radius.  Numerical ray tracing verifies these qualitative trends in the apparent shadow. Using the 6$^{th}$ - order WKB approximation method, we also calculate the scalar quasinormal modes and determine that oscillation frequencies and damping rates behave oppositely according to the sign and magnitude of each parameter. By comparing the shadow radius with the recent Event Horizon Telescope constraints on the Sgr A*, we find a narrow window of parameter space that agrees with observed data. In other words, the coupling parameters should be very small.  These measurements restrict modified‑gravity impacts within realistic astrophysical contexts.

\end{abstract}

\maketitle

\section{Introduction}

General Relativity (GR) has stood for over a century as the cornerstone of our understanding of gravitation, accurately predicting phenomena across an impressive range of scales—from the perihelion precession of Mercury to the most recent detections of gravitational waves by LIGO and Virgo \cite{Abbott2009Jun,TheLIGOScientificCollaboration2017Oct,LIGOScientificCollaborationandVirgoCollaboration2017Oct}. However, persistent puzzles on both the largest and smallest scales suggest that GR may not be the final word in gravitational physics. These puzzles include various observational aspects like the anomalous behaviour of rotational velocity curves of galaxies, energy content of the Universe and so on. These considerations have motivated the exploration of modified gravity theories that extend GR by introducing additional degrees of freedom. Among the simplest and most theoretically well-motivated of these extensions are scalar–tensor theories, in which a scalar field $\phi$ couples non-minimally to the metric. Lovelock’s theorem demonstrates that in four dimensions any diffeomorphism-invariant action built solely from the metric and yielding second-order field equations must coincide with the Einstein–Hilbert action (possibly plus a cosmological constant) \cite{Lovelock1971Mar}. Consequently, any modification involving only the metric inevitably introduces additional fields.
Pure Horndeski theories, by construction, maintain second‐order field equations and thus avoid the need for delicate degeneracy conditions to eliminate Ostrogradsky ghosts.  This simplicity allows for fully analytic black hole (BH) solutions such as the static, asymptotically flat solution of \cite{Babichev2024Jul}--and renders the computation of thermodynamic quantities (mass, temperature, entropy) straightforward and transparent. Moreover, pure Horndeski BHs exhibit clear quasi‐normal mode spectra and tidal Love numbers that deviate from General Relativity only at well-controlled orders \cite{Minamitsuji2022May,Bakopoulos2024Jul}. Consequently, pure Horndeski offers a minimal, mathematically robust framework for exploring scalar–tensor BH physics. It has recently been demonstrated that both Horndeski and beyond-Horndeski frameworks admit black-hole configurations characterized by a static, non-Schwarzschild geometry accompanied by a non-trivial scalar-field profile \cite{Baake2024Mar,Bakopoulos2024Jan,Babichev2017Apr}. These departures from the standard general-relativistic solutions present exciting opportunities for observational tests in the near future.

Over the past decade, significant progress has been made in constructing and analyzing Horndeski BHs. The recent study in pure Horndeski theory presented rotating solutions with non-analytic couplings leading to deformations of Kerr and novel ergoregions \cite{Walia2022Jun}. Complementary work on spontaneous scalarization via a Gauss–Bonnet coupling demonstrated that BHs acquire scalar hair above a critical coupling threshold, indicating bifurcations in the solution space \cite{Doneva2018Mar}. Linear stability analyses have since computed quasi-normal mode spectra for both static and rotating solutions, confirming dynamical viability regions \cite{Minamitsuji2022May}. In addition, deviations in tidal Love numbers relative to GR predictions were quantified, pointing to potential observable imprints in gravitational-wave ringdown signals \cite{Bakopoulos2024Jul}.
Numerical studies through backward ray tracing have explored optical signatures of scalar hair, mapping out BH shadow contours that deviate from Kerr expectations \cite{Cunha2015Nov}. Foundational works on stealth rotating solutions in shift-symmetric Horndeski sectors and on cosmological BH embeddings have broadened the catalog of known solutions \cite{Bakopoulos2025Jan,Charmousis2019Oct}. Meanwhile, investigations into wormhole and exotic compact object configurations within generalized scalar–tensor frameworks underscore the versatility of Horndeski interactions in admitting a wide variety of horizonless and horizon-full spacetimes \cite{Mironov2019Jun,Chatzifotis2022Mar,Vlachos2021Feb,Brihaye2020Dec}.
Looking ahead, a systematic understanding of scalarized BH solutions in Horndeski and DHOST theories is crucial for confronting gravitational models with current and future observations. Gravitational-wave detectors (e.g., LISA, Einstein Telescope) and very-long-baseline interferometry (Event Horizon Telescope) promise unprecedented probes of the strong-field regime, where deviations from Kerr might manifest as modifications to ringdown frequencies, BH shadows, or tidal Love numbers \cite{Cardoso2017Apr,Diedrichs2025Jan}.  At the same time, comprehensive thermodynamic analyses will illuminate the stability and phase structure of these hairy BHs, potentially revealing new transitions or instabilities absent in GR \cite{Walia2022Jun,Sekhmani2025Mar,Hennigar2015Dec,Zou2014Nov,Cavalcanti2022Jul}.

To the best of our knowledge, the thermodynamical, optical and QNM ringing properties have not yet been explored for a pure Horndeski BH system embedded in a PFDM surrounding. Therefore, we plan this paper in the following way: 
In section \ref{sec2}, we first review the theoretical foundations of Horndeski theories and the formulation of the Horndeski BH in the ambience of a PFDM. Next, in section \ref{sec3}, we discuss the thermodynamical analysis of our model and how the Horndeski and PFDM play a role in the thermodynamical stability of the BH. In section \ref{sec4}, we employ the backward ray-tracing technique to solve the null-geodesic equations.  In section \ref{sec5}, we calculate the shadow radius and confront the same with the observational shadow data to obtain bounds on the Horndeski and PFDM parameters. In section \ref{sec6}, we calculate the Quasinormal modes of the BH system using the 6$^{th}$-order WKB method. Finally, in section \ref{sec7} we summarize our work with the conclusion.

\section{Quartic Square-Root Horndeski BH in the Field of Perfect Fluid Dark Matter}
\label{sec2}
A particularly rich subclass of scalar–tensor theories is encapsulated by the Horndeski action, first formulated in 1974. Horndeski theories represent the most general four-dimensional scalar–tensor Lagrangians whose field equations remain second-order, thereby avoiding the Ostrogradsky instability associated with higher derivatives \cite{Kobayashi2019Jul}. In modern notation, the Horndeski action (in pure case) is represented as \cite{Babichev2024Jul}
\begin{equation}
S = \int d^4 x \sqrt{-g} ( \mathcal{L}_2 + \mathcal{L}_3  + \mathcal{L}_4  + \mathcal{L}_5 )
\label{action}
\end{equation}
The Lagrangian can be expressed in terms of four arbitrary functions $G_2(X),\dots,G_5(X)$, where $X\equiv-\tfrac12\nabla_\mu\phi\nabla^\mu\phi$, as

\begin{equation}\label{horndeski_lag}
	\begin{aligned}
		\mathcal{L}_2 &= G_2(X)\,, \\
		\mathcal{L}_3 &= -G_3(X)\,\Box\phi\,, \\
		\mathcal{L}_4 &= G_4(X)\,R + G_{4,X}(X)\bigl[(\Box\phi)^2 - \phi_{\mu\nu}\phi^{\mu\nu}\bigr]\,, \\
		\mathcal{L}_5 &= G_5(X)\,G_{\mu\nu}\phi^{\mu\nu}
		- \tfrac{1}{6}G_{5,X}(X)\bigl[(\Box\phi)^3 - 3\Box\phi\,\phi_{\mu\nu}\phi^{\mu\nu}
		\\&\hspace{4cm}+ 2\phi_{\mu\nu}\phi^{\nu\rho}\phi_{\rho}{}^{\mu}\bigr]
	\end{aligned}
\end{equation}

Here $G_{i,X}\equiv\partial G_i/\partial X$ and $\phi_{\mu\nu}\equiv\nabla_\mu\nabla_\nu\phi$ \cite{Babichev2017Apr,Babichev2024Jul}.

The quartic Horndeski model with a square-root kinetic coupling represents a particularly elegant sector of scalar–tensor gravity. It falls within the original Horndeski class and maintains second-order field equations. Specifically, we consider the Lagrangian \cite{Babichev2024Jul,Babichev2017Apr}
\begin{align}
	G_2 &= \eta X, \nonumber\\
	G_4 &= 1 + \beta \sqrt{-X}, \nonumber\\
	G_3 &= G_5 = 0, \label{eq:nanalyticG}
\end{align}
where $\eta$ and $\beta$. This theory includes a canonical kinetic term, a standard Einstein–Hilbert piece represented by 1 in $G_4$ \cite{Babichev2024Jul}, and a non-analytic square-root term that introduces scalar–tensor mixing. Thus the corresponding action reads \cite{Babichev2017Apr}:
\begin{equation}
	\begin{split}
		S = \int d^4x\,\sqrt{-g}\Bigl\{ 
		\bigl[1 + \beta\,\sqrt{(\partial\phi)^2/2}\bigr]\,R
		- \tfrac{\eta}{2}\,(\partial\phi)^2 \\[6pt]
		-\,\frac{\beta}{\sqrt{2(\partial\phi)^2}}\bigl[(\Box\phi)^2 - (\nabla_\mu\nabla_\nu\phi)^2\bigr]
		\Bigr\}\,.
	\end{split}
	\label{eq:quartic_sqrt_action}
\end{equation}

Using the static, spherically symmetric ansatz for the metric
\begin{equation}
	ds^2 = -f(r)dt^2 + \frac{dr^2}{f(r)} + r^2 d\Omega^2,
	\label{eq:ansatz}
\end{equation}

Substituting into the Einstein equations gives the solution \cite{Babichev2024Jul}
\begin{equation}
	f(r) = 1 - \frac{2M}{r} - \frac{\beta^2}{2 \eta r^2},
	\label{eq:f_metric}
\end{equation}
where $\beta$ and $\eta$ are the two parameters of the model and $M$ is the ADM mass of the spacetime. The existence of a non‐trivial solution requires that the coupling constants $\beta$ and $\eta$ share the same sign \cite{Babichev2024Jul}. In the limit $\beta \to 0$, the metric smoothly reduces to the trivial Schwarzschild geometry. At spatial infinity, the spacetime is asymptotically flat. Also, a curvature singularity persists at $r=0$. For $\eta>0$, the geometry exhibits a single horizon. Conversely, when $\eta<0$, the condition
\(M < \frac{|\beta|}{\sqrt{-2\eta}} \) there exists no horizon, yielding a naked singularity, whereas
\(M > \frac{|\beta|}{\sqrt{-2\eta}}\)
admits two distinct horizons \cite{Babichev2017Apr,Babichev2024Jul}.

In this work, we want to explore the interesting aspect of the quartic square-root Horndeski BH when it is surrounded by perfect fluid dark matter (PFDM). 
Observational evidence from the galactic rotation curves and gravitational lensing has hinted that an elusive form of dark matter permeates the halos of most galaxies, implying that any realistic astrophysical BH can be immersed in a non-negligible dark matter distribution.  By modelling this dark matter component as a perfect fluid, one obtains an exact, analytic BH solution parametrized by the PFDM constant $\alpha$, which encodes the asymptotic density profile and corresponding gravitational backreaction of the dark matter halo \cite{Xu2018Apr}.  This framework allows us to quantify, in closed form, how dark matter alters the spacetime geometry near the horizon, thereby modifying physical properties such as the photon‐sphere radius, circular orbit structure, shadows and QNM frequencies.  Given the updated resolution of instruments like the EHT, even a subtle PFDM-induced effect on the BH shadow or lensing rings could become measurable. This could provide a promising observational probe of the local dark matter environment.

 In order to introduce the effects of PFDM, we may consider the energy momentum tensor of PFDM given by $T_{\mu}^{\nu, PFDM} = \text{diag} [-\rho , p_r, p_\theta, p_\phi]$ \cite{Xu2018Apr}, where $\rho$ is the energy density of the PFDM and $p_r$, $p_\theta$ and $p_\phi$ are the radial and tangential pressures. As referred in \cite{Das2022Mar}, the energy density and the pressures can be expressed as:
\begin{equation}
	\rho = - p_r = \frac{b}{8\pi r^3} \quad \text{and} \quad p_\theta = p_\phi = - \frac{b}{8\pi r^3}.
	\label{pfdm_rho_p} 
\end{equation}
It turns out that, for a spherically symmetric spacetime, the metric lapse function modifies into \cite{Xu2018Apr}
\begin{equation}
		f(r) = 1 - \frac{2M}{r} - \frac{\beta^2}{2 \eta r^2} + \frac{b}{r} \ln \left( \frac{r}{b}\right)
		\label{eq:f_metric_pfdm}
	\end{equation}
where, $b$ is the model parameter associated the effect of the PFDM. In the vanishing limit of $b \to 0$, the original metric \eqref{eq:f_metric} is recovered.

\begin{figure*}[htb]
	\centerline{\includegraphics[scale = 0.4]{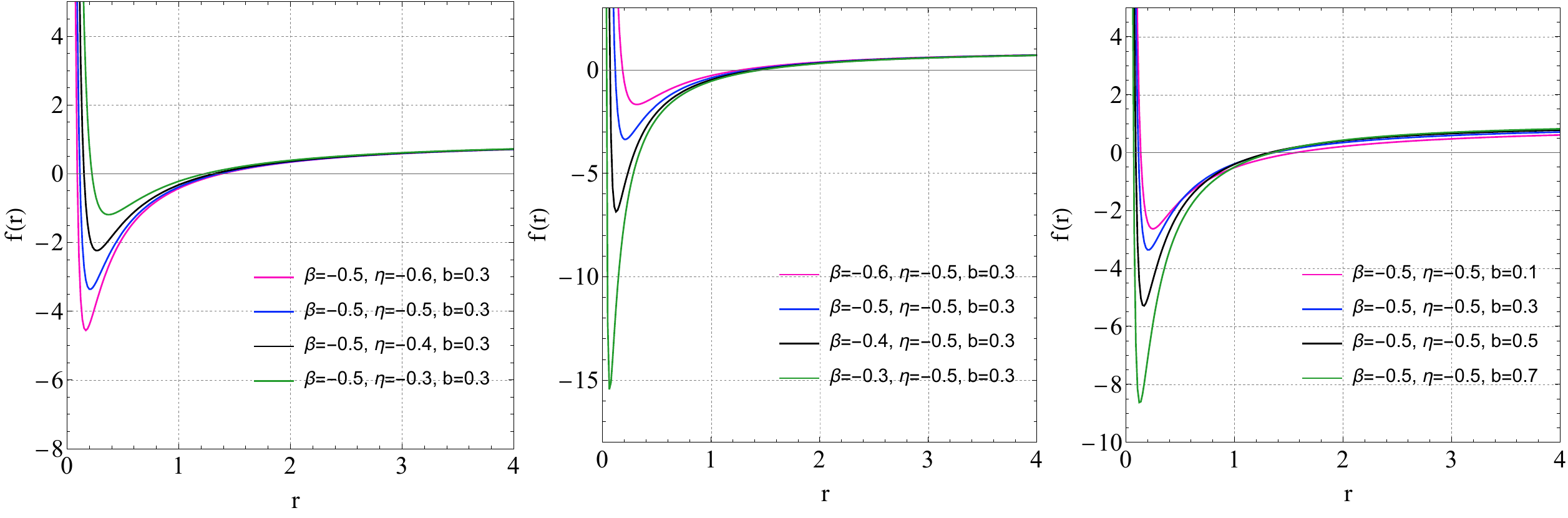}}
	\caption{The lapse function is plotted for different combinations of the parameters $\beta$, $\eta$ and $b$.}
	\label{lapse_plot}
\end{figure*}
Fig. \ref{lapse_plot} depicts the lapse function of the BH with respect to the Horndeski parameters and the PFDM parameters.
The lapse function exhibits three characteristic behaviours as each parameter is varied. In the leftmost panel, increasing \(\eta\) from \(-0.6\) to \(-0.3\) (with \(\beta=-0.5\), \(b=0.3\)) uniformly raises the curve at intermediate radii and shifts the horizon radius \(r_h\) (where \(f(r_h)=0\)) to smaller values.  In the middle panel, decreasing \(\lvert\beta\rvert\) (from \(-0.6\) to \(-0.3\)) similarly reduces the depth of the  trough and shifts \(r_h\) inwards.  Finally, in the rightmost panel, increasing the PFDM parameter \(b\) from \(0.1\) to \(0.7\)  and shift the horizon towards smaller values.  In all cases, there exist two horizons, and each parameter primarily governs the location of the horizons and hence the effective size of the BH. Moreover, the solution is asymptotically flat since $f(r) \to 1$ as $r \to \infty$.

\section{Thermodynamical Analysis}
\label{sec3}
In this section, let us calculate the thermodynamic functions of the Horndeski-PFDM BH system. We begin by calculating the Hawking temperature $T_h$, through the direct relation with the metric function  Eq. \eqref{eq:f_metric_pfdm}, which is 
\begin{equation}
T = \frac{f'(r)}{4\pi} = \frac{\beta ^2+2 \eta  r_h \left(b+r_h\right)}{8 \pi  \eta  r_h^3},
\label{temp}
\end{equation}
where $r_h$ is the horizon radius.
\begin{figure*}[htb]
\centerline{\includegraphics[scale=0.45]{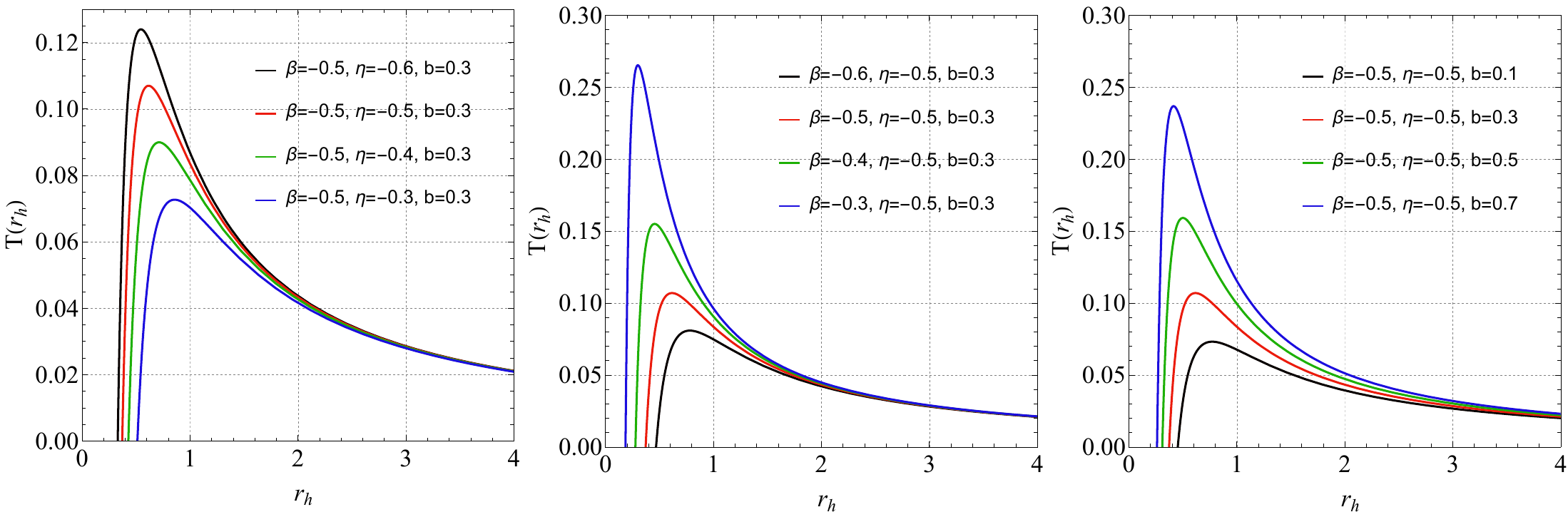}}
\caption{The temperature is plotted for various values of the Horndeski parameters and the PFDM parameter.}
\label{temp_plot}
\end{figure*}

Fig.~\ref{temp_plot} shows the Hawking temperature \(T\) as a function of the horizon radius \(r_{h}\), illustrating its dependence on the model parameters \(\eta\), \(\beta\), and the PFDM parameter \(b\). As \(r_{h}\) increases from its minimal value at which \(T = 0\), the temperature initially rises, attains a maximum, and subsequently decreases for larger horizon radii. The existence of such a peak in the temperature profile reflects a characteristic thermodynamic behaviour commonly associated with BHs in extended phase space, often indicating the presence of multiple thermodynamic branches and a possible phase transition. This local maximum corresponds to a divergence of the heat capacity, signalling a transition from a thermodynamically stable to an unstable BH branch. Holding \(\beta\) and \(b\) fixed, the maximum temperature monotonically decreases as \(\eta\) increases. Conversely, for fixed \(\eta\) and \(b\), the peak value of \(T\) grows with increasing \(\beta\). When \(\eta\) and \(\beta\) are held constant, an increase in \(b\) likewise elevates the maximum of the temperature curve. In the limit of sufficiently large \(r_{h}\), \(T\) asymptotically approaches zero. Below the minimum horizon radius, referring to \(T = 0\), the temperature becomes negative. This indicates the absence of physically admissible BH states.

Again, setting $f(r_h) = 0$ leads us to obtain the Arnowitt-Deser-Misner (ADM) mass ($M$) of the BH given by
\begin{equation}
M = \frac{-\beta ^2+2 b \eta  r_h \log \left(\frac{r_h}{b}\right)+2 \eta  r_h^2}{4 \eta  r_h},
\label{mass}
\end{equation}
Clearly, in the limiting values of $\beta, \, b \, \to 0$, the mass of Schwarszchild BH is recovered,
\begin{equation}
M_{sch} = \frac{r_h}{2},
\label{sch_mass} 
\end{equation}
\begin{figure*}
\centerline{\includegraphics[scale=0.45]{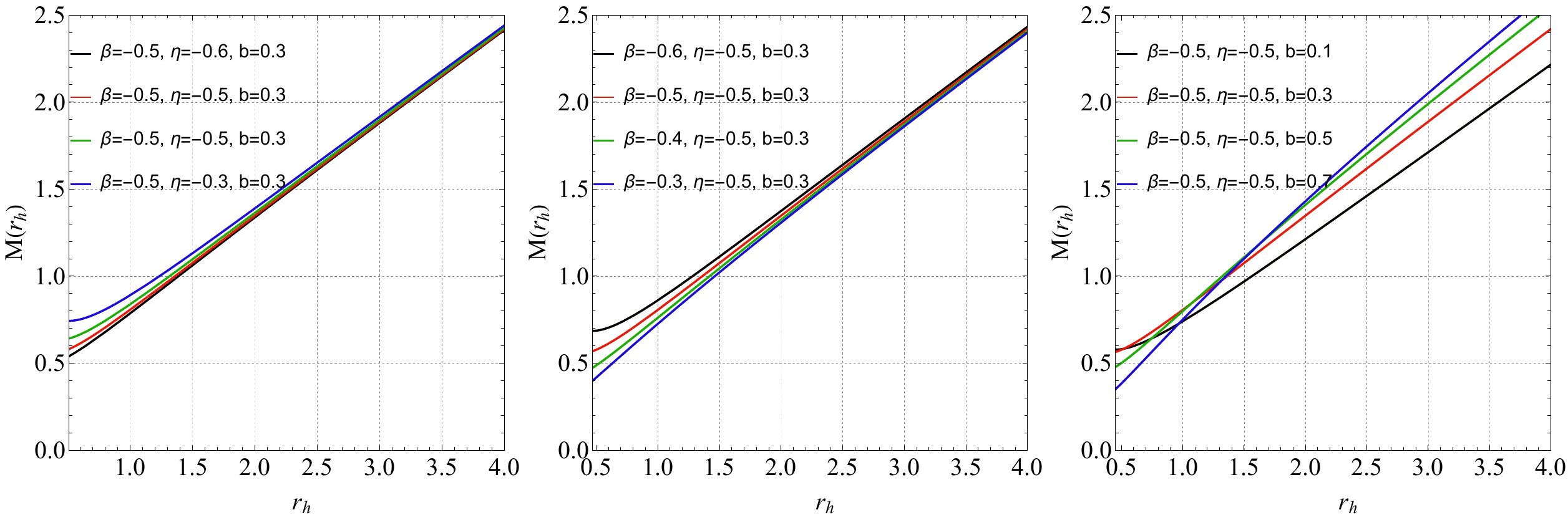}}
\caption{The mass is plotted with respect to horizon radius for different combinations of the parameters $\eta$, $\beta$ and $b$.}
\label{mass_plot}
\end{figure*}
From Fig. \ref{mass_plot} we see that the ADM mass increases monotonically with increasing horizon radius for all the mentioned values of the parameters in the plot. 

\begin{figure*}
\centerline{\includegraphics[scale=0.45]{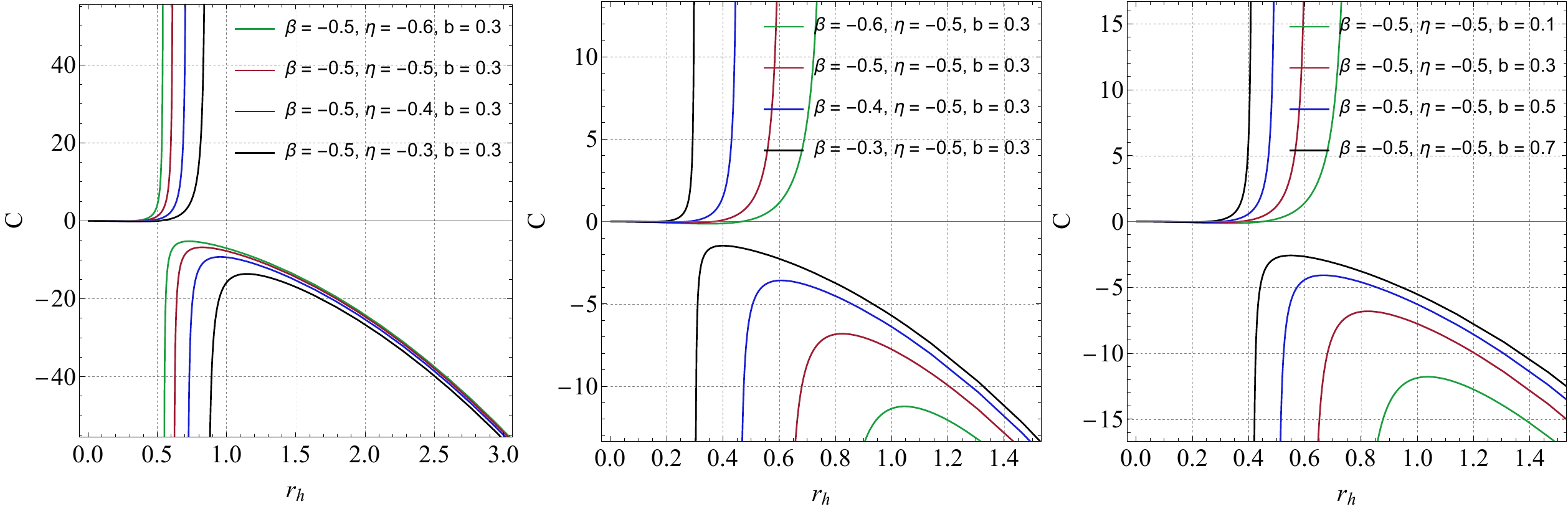}}
\caption{The specific heat is plotted for different values combinations of $\eta$, $\beta$ and $b$.}
\label{spheat_plot}
\end{figure*}

It is observed from Fig.~\ref{mass_plot} that, when restricted to the physically admissible branch for which \(T(r_{h})>0\), the mass function \(M(r_{h})\) exhibits a strictly monotonic increase with respect to the horizon radius \(r_{h}\).  Specifically,
\[
\frac{dM}{dr_{h}} > 0 
\quad \text{for all } r_{h} \text{ such that } T(r_{h})>0,
\]
implying a one‐to‐one correspondence between the ADM mass and \(r_{h}\) in the positive‐temperature regime.  Consequently, no additional extrema appear in \(M(r_{h})\) that could indicate further thermodynamic instabilities within this domain.
In our BH system, we obtain the specific heat as:
\begin{equation}
	C = \frac{\partial M/ \partial r_{h}}{\partial T/ \partial r_{h}} =  -\frac{2 \pi  r^2 \left(\beta ^2+2 \eta  r_h (b+ r_h)\right)}{3 \beta ^2+2 \eta  r_h (2 b+r_h)},
	\label{spheat}
\end{equation}
 From the relation
\[
C(r_{h}) = \frac{\partial M/dr_{h}}{\partial T/dr_{h}},
\]
it follows that at the peak radius \(r_{\rm peak}\), where \(\bigl.\tfrac{\partial T}{\partial r_{h}}\bigr|_{r_{\rm peak}} = 0\), the specific heat \(C\) satisfies
\[
C \;\longrightarrow\; 
\begin{cases}
+\infty, & r_{h} \to r_{\rm peak}^{-},\\
-\infty, & r_{h} \to r_{\rm peak}^{+}.
\end{cases}
\]
Fig.~\ref{spheat_plot} displays the corresponding specific‐heat curves, which diverge precisely at those radii where the temperature curves (Fig.~\ref{temp_plot}) attain local maxima. 
For \(r_{h} < r_{\rm peak}\), \(\partial T/\partial r_{h} > 0\) and thus \(C>0\), indicating thermodynamic stability; for \(r_{h} > r_{\rm peak}\), \(\partial T/\partial r_{h} < 0\) and hence \(C<0\), indicating instability.  Moreover, variations of the model parameters (\(\eta\), \(\beta\), or \(b\)) shift both the location and magnitude of \(T_{\rm peak}\), and the corresponding divergence in \(C(r_{h})\) shifts accordingly.  In particular, an increase in \(\eta\) (with \(\beta\) and \(b\) held fixed) lowers and shifts the temperature maximum to smaller \(r_{h}\), causing the vertical asymptote of \(C\) to move inward.  In contrast, increasing either \(\beta\) or \(b\) (with the other parameters fixed) raises and shifts \(T_{\rm peak}\) to larger \(r_{h}\), and the \(C\)‐divergence follows this displacement.

By collecting the information from Figs.~\ref{mass_plot}, \ref{temp_plot} and \ref{spheat_plot}, one identifies three distinct regimes of black‐hole solutions.  Firstly, for \(r_{h} < r_{\rm peak}\), the temperature \(T(r_{h})\) increases with \(r_{h}\) (implying \(C>0\)) while \(M(r_{h})\) also increases; this branch represents locally thermodynamically stable BHs that exist for \(T < T_{\rm peak}\).  Secondly, for \(r_{h} > r_{\rm peak}\), \(T(r_{h})\) decreases with increasing \(r_{h}\) (hence \(C<0\)), although \(M(r_{h})\) continues to grow.  Such configurations are thermodynamically unstable and will tend to evolve back toward the stability boundary at \(r_{\rm peak}\) under small perturbations.  Thirdly, below the minimal radius \(r_{\rm min}\) (where \(T(r_{\rm min}) = 0\)), no physical black‐hole solutions exist since \(T<0\).  An evaporating BH will follow the stable branch down to \(r_{h} = r_{\rm peak}\), linger in a quasi‐stable regime (where \(\lvert C\rvert \to \infty\)), and ultimately approach the non‐evaporating remnant at \(r_{h} = r_{\rm min}\).

We can calculate the entropy of the BH as
\begin{equation}
S_{BH} = \int \frac{dM}{T} = \pi r_h^2 / 4,
\label{bh_entropy}
\end{equation}
which shows that the BH follows the usual Bekenstein-Hawking entropy.

In order to infer the global stability of the Horndeski BH system immersed in the PFDM, we can analyze the Helmholtz free energy ($F$) for the BH system.The thermodynamic system is said to be globally unstable if $F$ is positive, and globally stable if $F$ is negative. Thus, the Helmholtz free energy can be found as
\begin{equation}
F = M - TS = \frac{1}{8} \left(4 b \ln \left(\frac{r_h}{b}\right)-2 b-\frac{3 \beta ^2}{\eta  r_h}+2 r_h \right)
\label{free_energy}
\end{equation}
\begin{figure*}
\centerline{\includegraphics[scale=0.45]{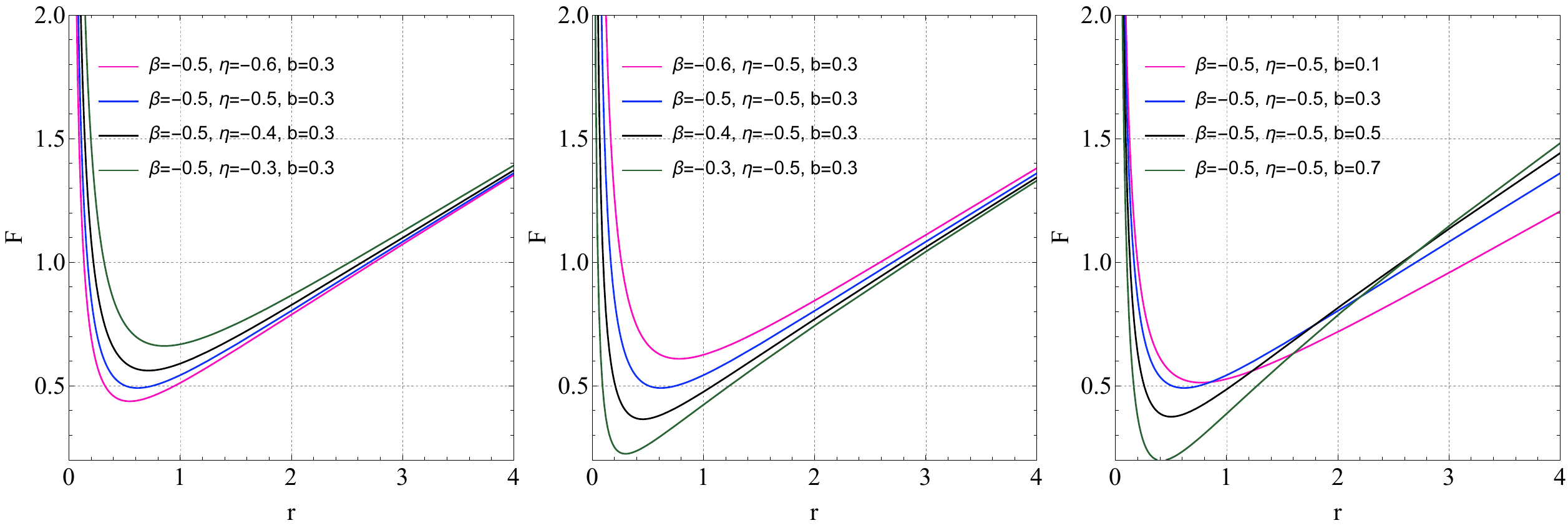}}
\caption{The free energy $F$ is shown for different values of $\eta$, $\beta$ and $b$.}
\label{free_energy_plot}
\end{figure*}

The global stability of BHs requires $F_{\mathrm{min}} < 0$ to favor the BH phase over the background spacetime. Increasing the Horndeski parameter $\eta$ (from $-0.6$ to $-0.3$) at fixed $\beta = -0.5$ and $b = 0.3$ reduces the stability by lowering $F_{\mathrm{min}}$. Also, when we increasw $\beta$ (from $-0.6$ to $-0.3$) at fixed $\eta = -0.5$ and $b = 0.3$,  $F_{\mathrm{min}}$ is reduced, decreases the instability. Larger PFDM parameter $b$ (from $0.1$ to $0.7$) at fixed $\beta$ and $\eta$ monotonically deepens $F_{\mathrm{min}}$, thus demonstrating the sensitive effect of dark matter parameter $b$ on the global stability. In all these cases, however, there is no transition into the negative $F$ domain. Thus, no Hawking-Page phase transition occurs. This behaviour is shown in Fig. \ref{free_energy_plot}.

\section{Geodesic Behaviour}
\label{sec4}
The paths of particles (massless or massive) in the gravitational field of a BH are determined by the functions \(A(r)\) and \(B(r)\), which act as a redshift factor and a measure of spatial curvature, respectively, in a spherically symmetric spacetime.  A conserved quantity along geodesics is produced by each of the two Killing vectors that are admitted in such a spacetime, reflecting invariance under temporal translations and rotations.  Specifically, the energy conservation generated by the timelike Killing vector is \[ E = A(r)\,\dot t,\]
Its motion is constrained to equatorial orbits (either elliptic, hyperbolic, or parabolic, depending on eccentricity) by the conserved angular momentum generated by the rotational Killing vector.  The formulation of the equations of motion and the derivation of an effective potential for both null and timelike trajectories, which determine orbital stability and precession, and these rely strongly on these conserved values. 

To derive the geodesic equations, we begin with the general form of the spherically symmetric line element,
\begin{equation}
ds^2 = - A(r)^2\,dt^2 + B(r)^2\,dr^2 + r^2\,d\theta^2 + r^2\sin^2\theta\,d\phi^2.
\label{gen_met}
\end{equation}
Because this metric is invariant under \(t\to t+\mathrm{const}\) and \(\phi\to\phi+\mathrm{const}\), any Killing vector \(K_\mu\) satisfies
\begin{equation}
K_\mu \,\dot x^\mu = \text{constant},
\label{kill}
\end{equation}
where \(\dot x^\mu = \frac{dx^\mu}{d\lambda}\) and \(\lambda\) is an affine parameter.  The explicit forms of the two relevant Killing vectors are
\begin{align}
K_\mu^{(t)} &= \bigl(-A(r),0,0,0\bigr), 
\label{time_killing}\\
K_\mu^{(\phi)} &= \bigl(0,0,0,r^2\sin^2\theta\bigr).
\label{sph_killing}
\end{align}
From these, one finds
\begin{equation}
E = A(r)\,\dot t,
\label{E-eqn}
\end{equation}
and, upon restricting motion to the equatorial plane \(\theta=\frac\pi2\),
\begin{equation}
L = r^2\,\dot\phi.
\label{L-eqn}
\end{equation}

The norm of the tangent vector,
\begin{equation}
\epsilon = -\,g_{\mu\nu}\,\dot x^\mu\,\dot x^\nu,
\label{eps}
\end{equation}
is also conserved (\(\epsilon=0\) for null geodesics, \(\epsilon=1\) for timelike geodesics).  Substituting the metric \eqref{gen_met} into \eqref{eps} gives
\begin{equation}
-\epsilon = -A(r)\,\dot t^2 + B(r)\,\dot r^2 + r^2\,\dot\phi^2,
\label{eps1}
\end{equation}
which can be rearranged to
\begin{equation}
\dot r^2 \;=\;\frac{E^2}{A(r)\,B(r)} \;-\;\frac{L^2}{r^2\,B(r)} \;-\;\frac{\epsilon}{B(r)}.
\label{rdotsq}
\end{equation}

Alternatively, one may start from the Lagrangian
\begin{equation}
\mathcal{L} = \frac{1}{2}\,g_{\mu\nu}\,\dot x^\mu\,\dot x^\nu
= \frac{1}{2}\bigl(-A(r)\,\dot t^2 + B(r)\,\dot r^2 + r^2\,\dot\phi^2\bigr),
\label{Lag}
\end{equation}
and apply the Euler–Lagrange equation in the \(r\)-direction,
\begin{equation}
\frac{d}{d\lambda}\Bigl(\frac{\partial\mathcal{L}}{\partial\dot r}\Bigr)
= \frac{\partial\mathcal{L}}{\partial r}.
\label{Lagr}
\end{equation}
With the conjugate momentum
\begin{equation}
p_r = \frac{\partial\mathcal{L}}{\partial\dot r} = B(r)\,\dot r,
\label{pr}
\end{equation}
one obtains
\begin{equation}
\dot p_r = \frac{1}{2}\Bigl(-A'(r)\,\dot t^2 + B'(r)\,\dot r^2 + 2r\,\dot\phi^2\Bigr).
\label{prdot}
\end{equation}

\subsection{Null Geodesics}
Using \eqref{E-eqn}, \eqref{L-eqn}, \eqref{pr} and \eqref{prdot}, the first-order system for null geodesics reads
\begin{equation}
\begin{aligned}
\dot t &= \frac{E}{A(r)},\\
\dot \phi &= \frac{L}{r^2},\\
\dot r &= \frac{p_r}{B(r)},\\
\dot p_r &= \frac{1}{2}\Bigl(-\frac{E^2}{A(r)^2}A'(r)
+ \frac{p_r^2}{B(r)^2}B'(r)
+ \frac{2L^2}{r^3}\Bigr).
\end{aligned}
\label{geod_eqs}
\end{equation}
Moreover, from \eqref{rdotsq} one may write
\begin{equation}
\tfrac12\,\dot r^2 + V_{\rm eff}(r) = \tfrac12\,E^2,
\label{eff1}
\end{equation}
with
\begin{equation}
V_{\rm eff}(r) = \frac{L^2}{2r^2\,B(r)} + \frac{\epsilon}{B(r)}.
\label{eff2}
\end{equation}

Utilizing the metric \eqref{eq:f_metric_pfdm}, and comparing with \eqref{gen_met}, we identify
\[
A(r) = f(r)\,, \qquad B(r) = \frac{1}{f(r)}\,.
\]
Hence the system \eqref{geod_eqs} becomes

\begin{equation}
\begin{aligned}
\dot{t} &= E \biggl(1 -\frac{2 M}{r} -\frac{b \ln \!\bigl(\frac{r}{b}\bigr)}{r}
-\frac{\beta ^2}{2 \eta  r^2} \biggr)^{-1},\\
\dot{\phi} &= \frac{L}{r^2},\\
\dot{r} &= p_r \,\biggl(1 -\frac{2 M}{r} -\frac{b \ln \!\bigl(\frac{r}{b}\bigr)}{r}
-\frac{\beta ^2}{2 \eta  r^2}\biggr),\\
\dot{p}_r &= -\frac{p_r^2 \bigl(\beta ^2-\eta  r (b-2 M)
+b \eta  r \ln \!\bigl(\frac{r}{b}\bigr)\bigr)}{2 \eta  r^3}
\\& -\frac{2 E^2 \eta  r \bigl(\beta ^2-\eta  r (b-2 M)
+b \eta  r \ln \!\bigl(\frac{r}{b}\bigr)\bigr)}
{\bigl(\beta ^2+2 b \eta  r \ln \!\bigl(\frac{r}{b}\bigr)
+2 \eta  r (2 M-r)\bigr)^2}
+\frac{L^2}{r^3}.
\end{aligned}
\label{geod_eqs_mod}
\end{equation}

\begin{figure*}[tbh]
	\centerline{\includegraphics[scale=0.5]{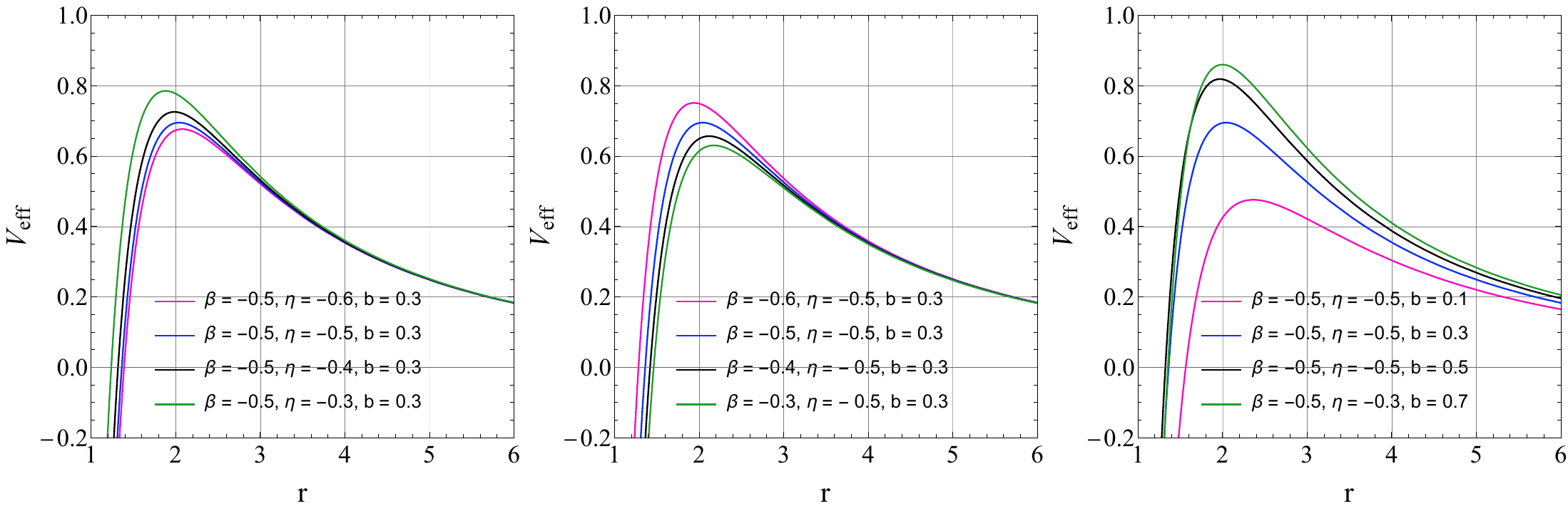}}
	\caption{The effective potential for null-geodesics is shown for various combinations of the model parameters.}
	\label{eff_pot_plot}
\end{figure*}

\begin{figure*}[tbh]
\centerline{\includegraphics[scale=0.4]{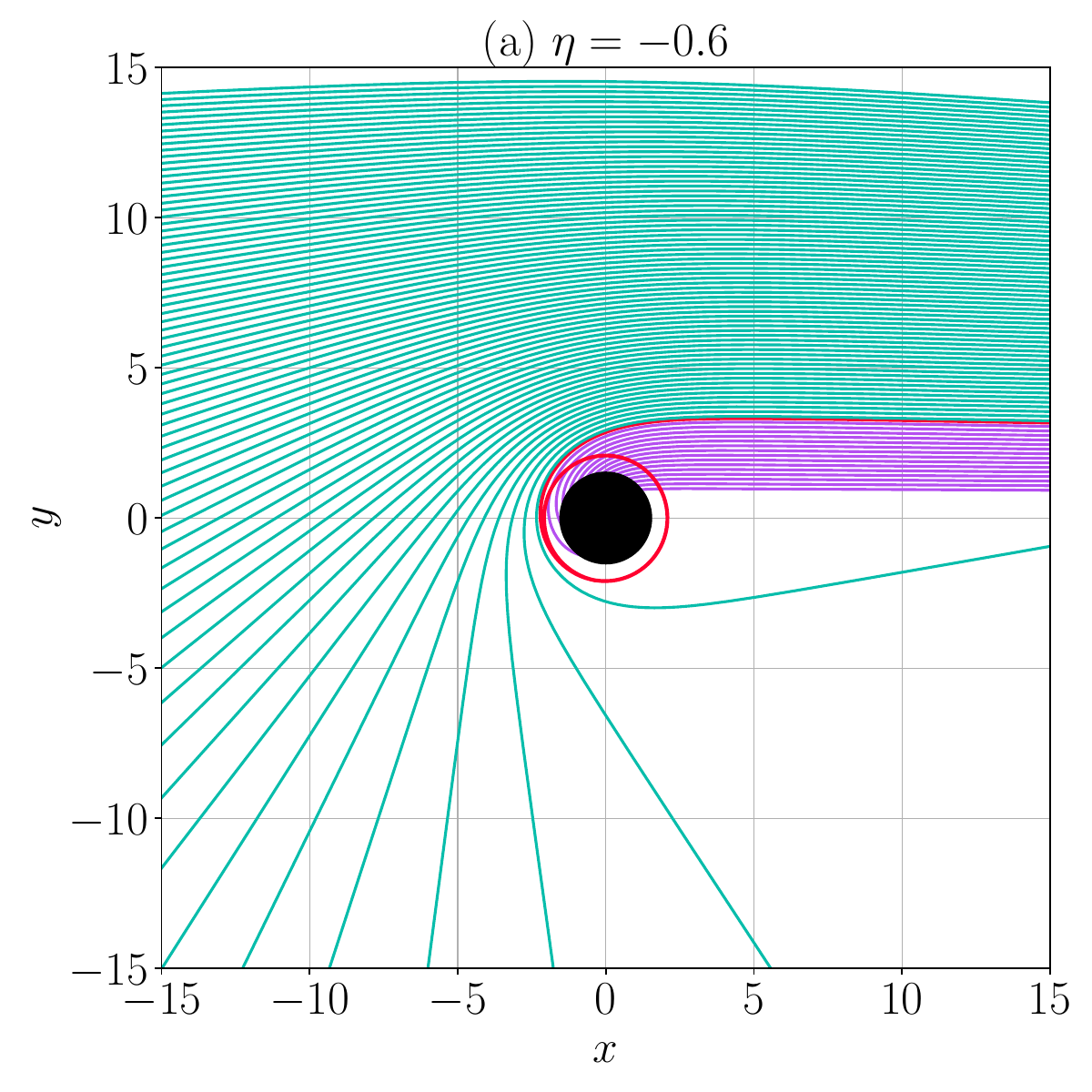}\hspace{-0.1cm}\includegraphics[scale=0.4]{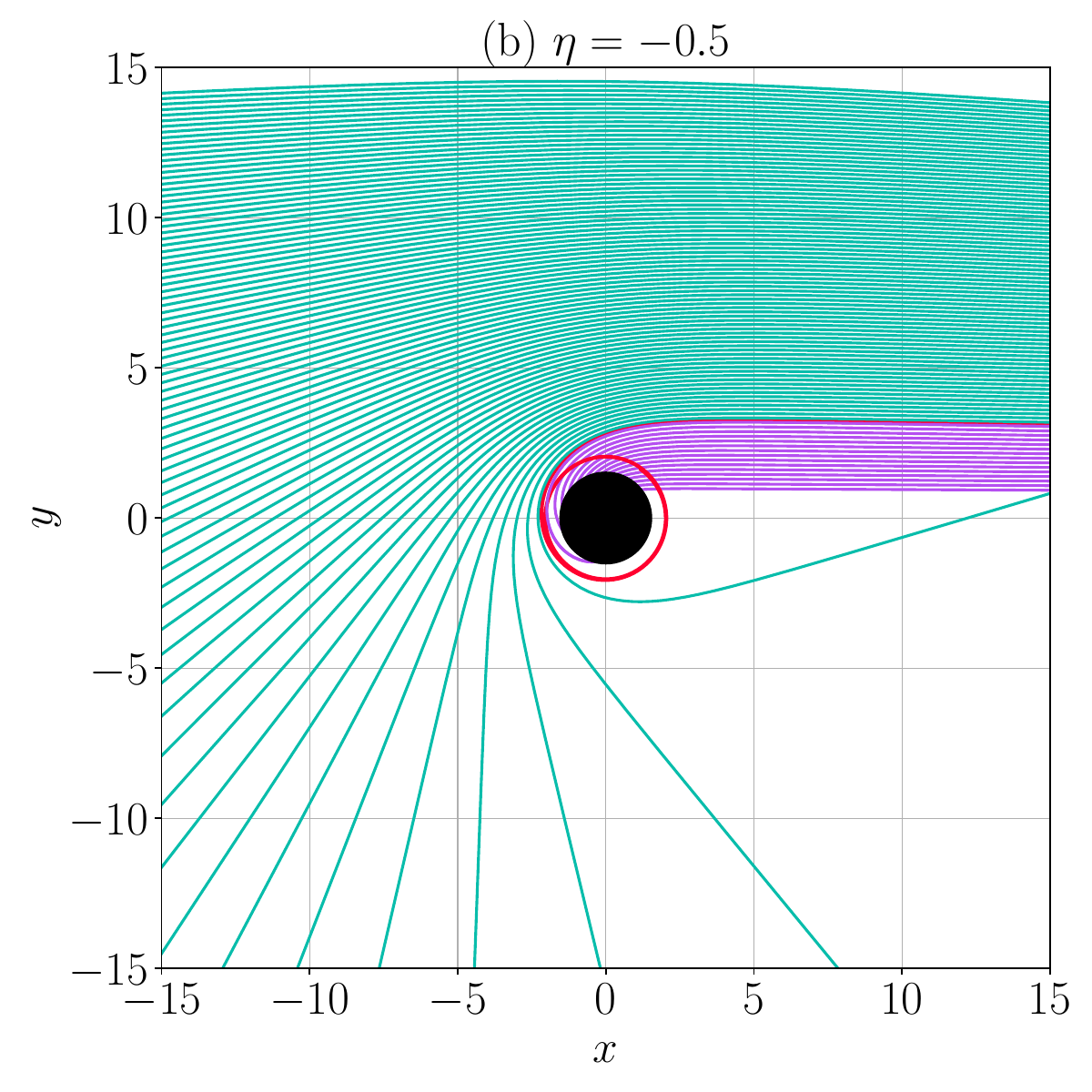}}
\centerline{\includegraphics[scale=0.4]{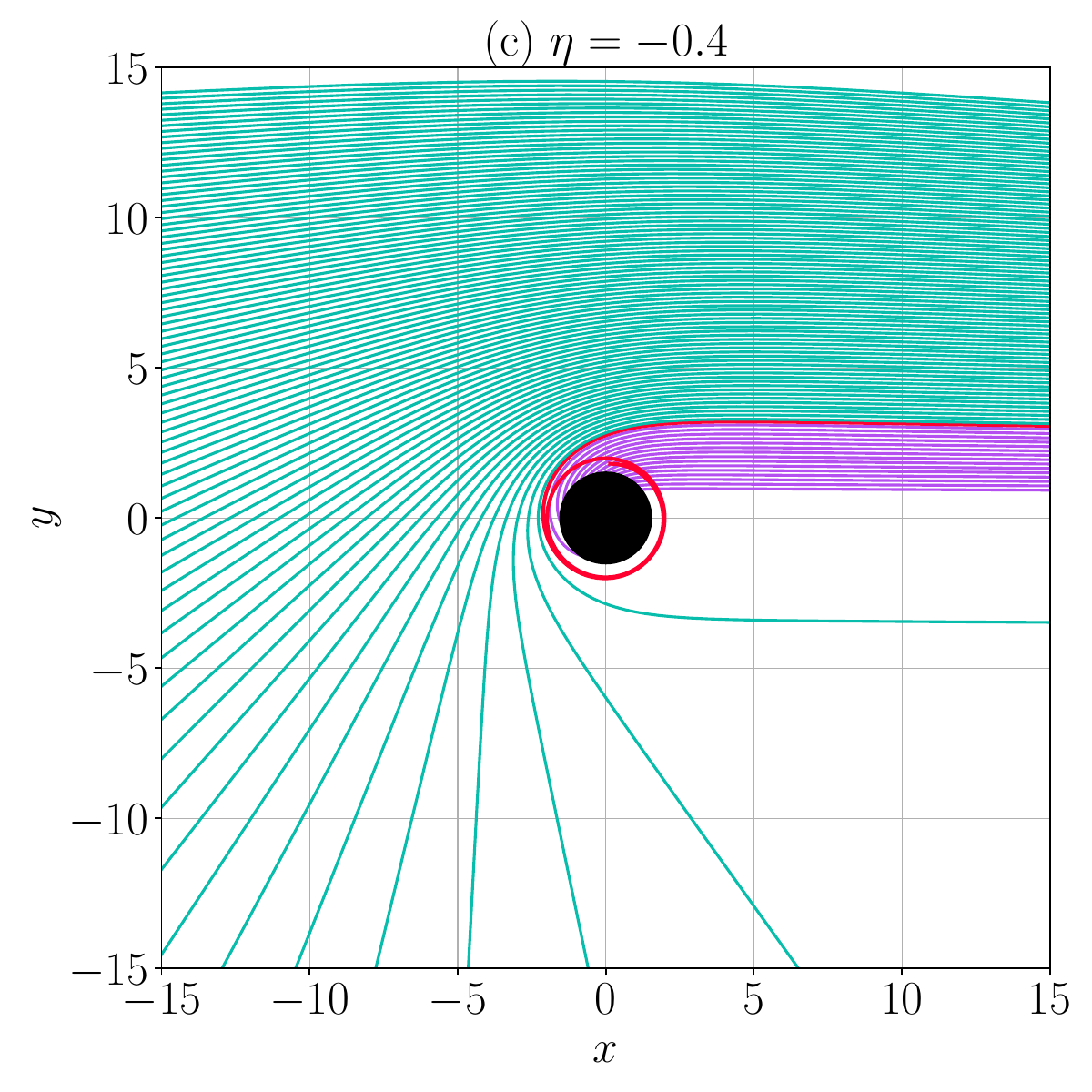}\hspace{-0.1cm}\includegraphics[scale=0.4]{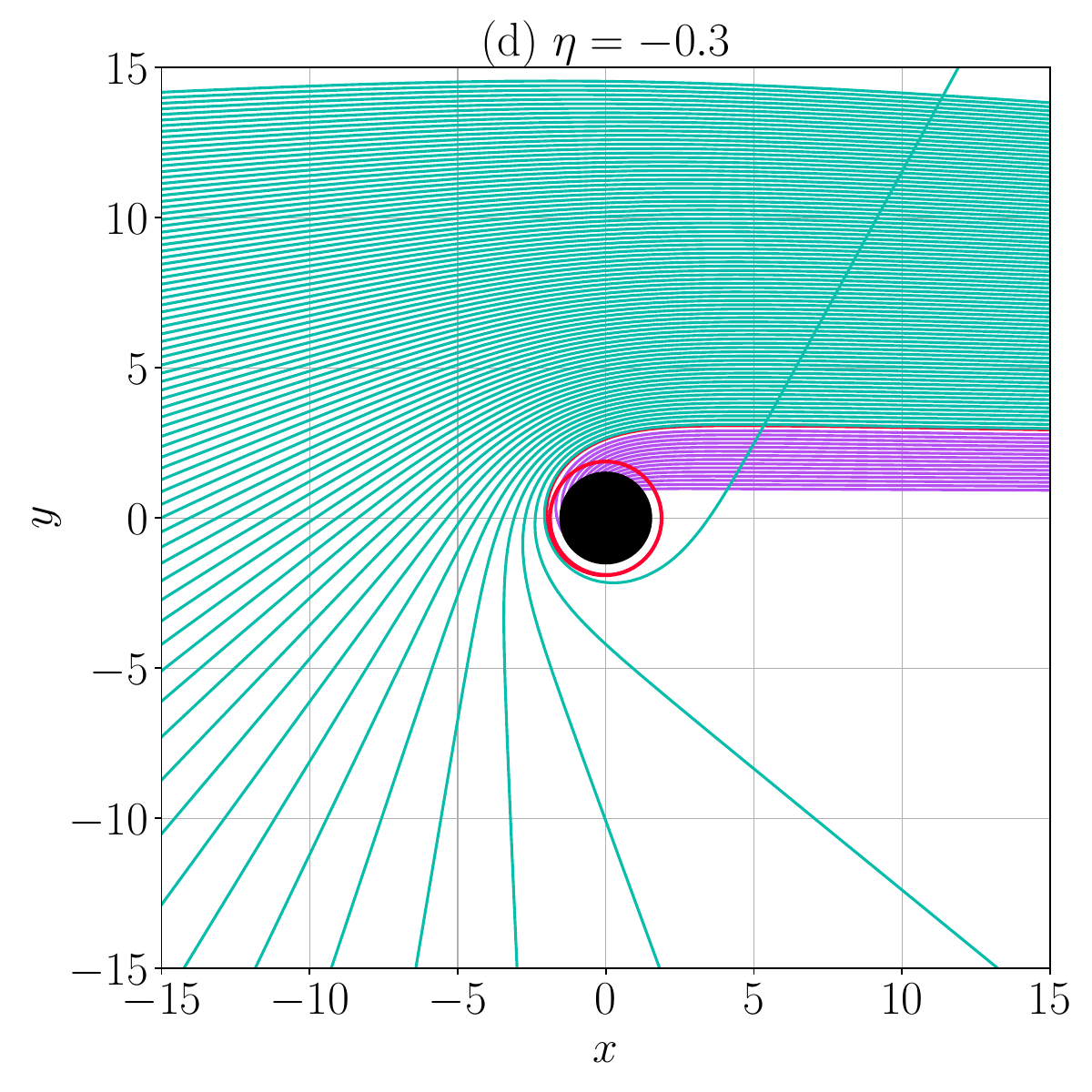}}
\caption{Geodesics of Null rays for different values of $\eta$ with fixed $\beta = -0.5$ and $b = 0.3$.}
\label{null_geod1}
\end{figure*}

\begin{figure*}[tbh]
	\centerline{\includegraphics[scale=0.4]{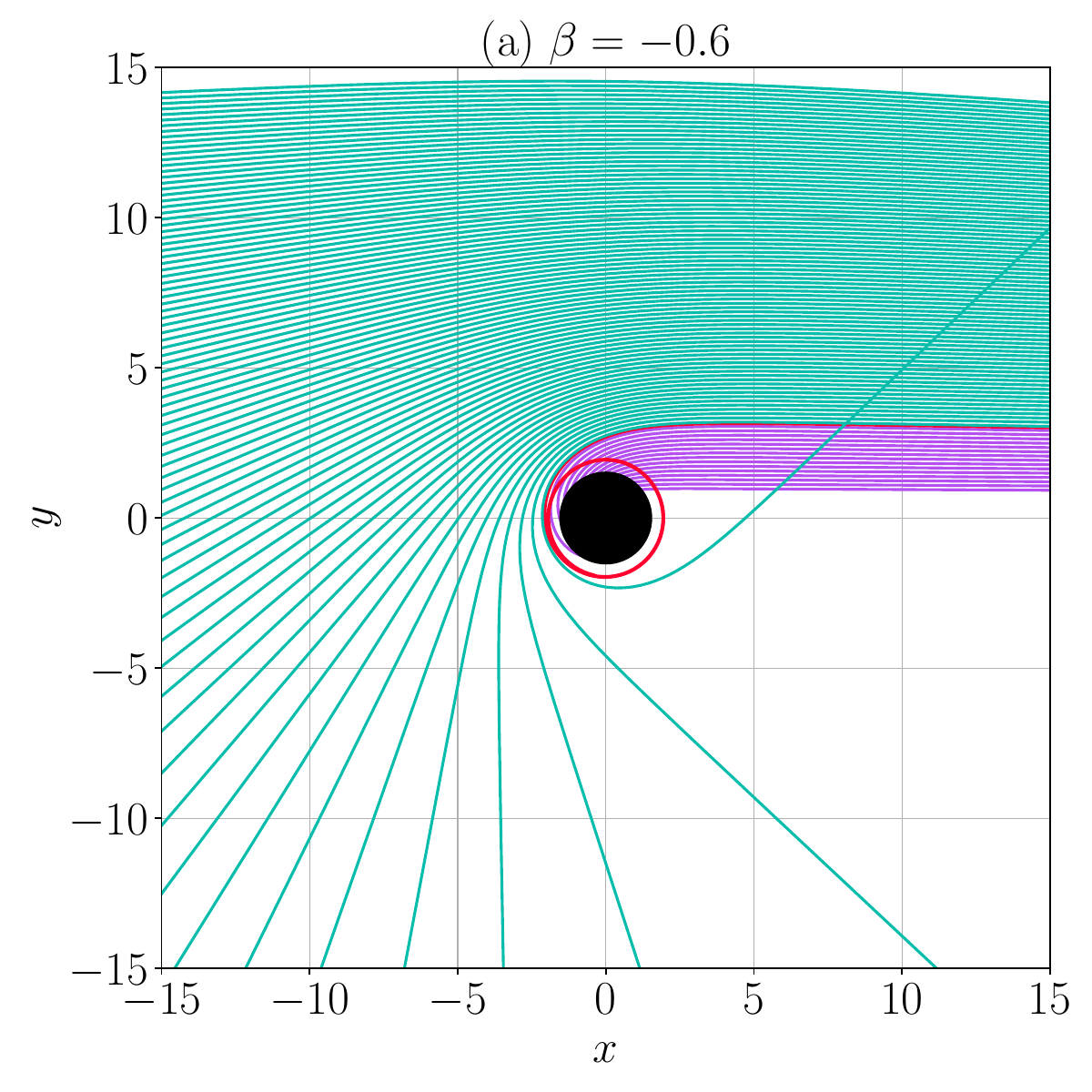}\hspace{-0.1cm}\includegraphics[scale=0.4]{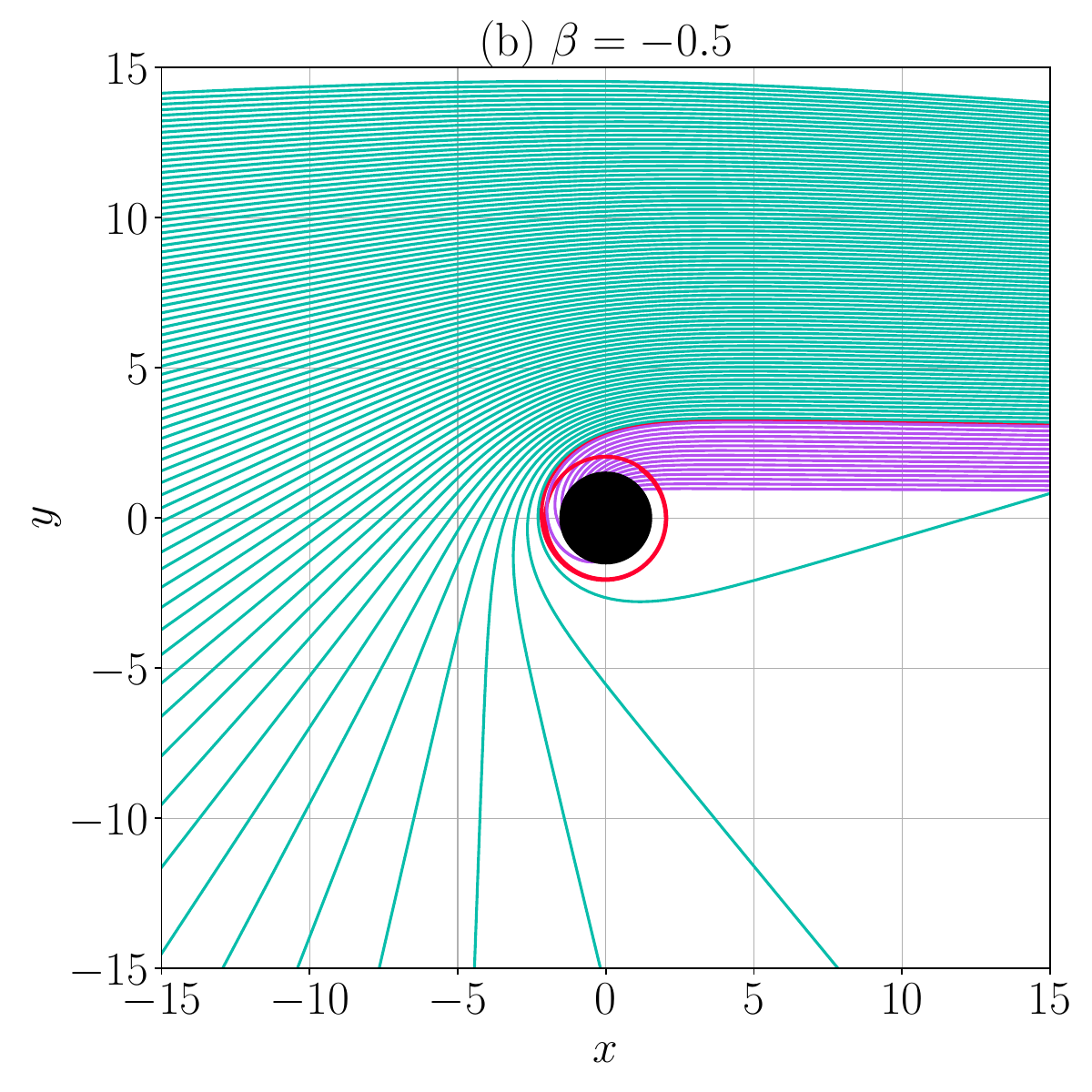}}
	\centerline{\includegraphics[scale=0.4]{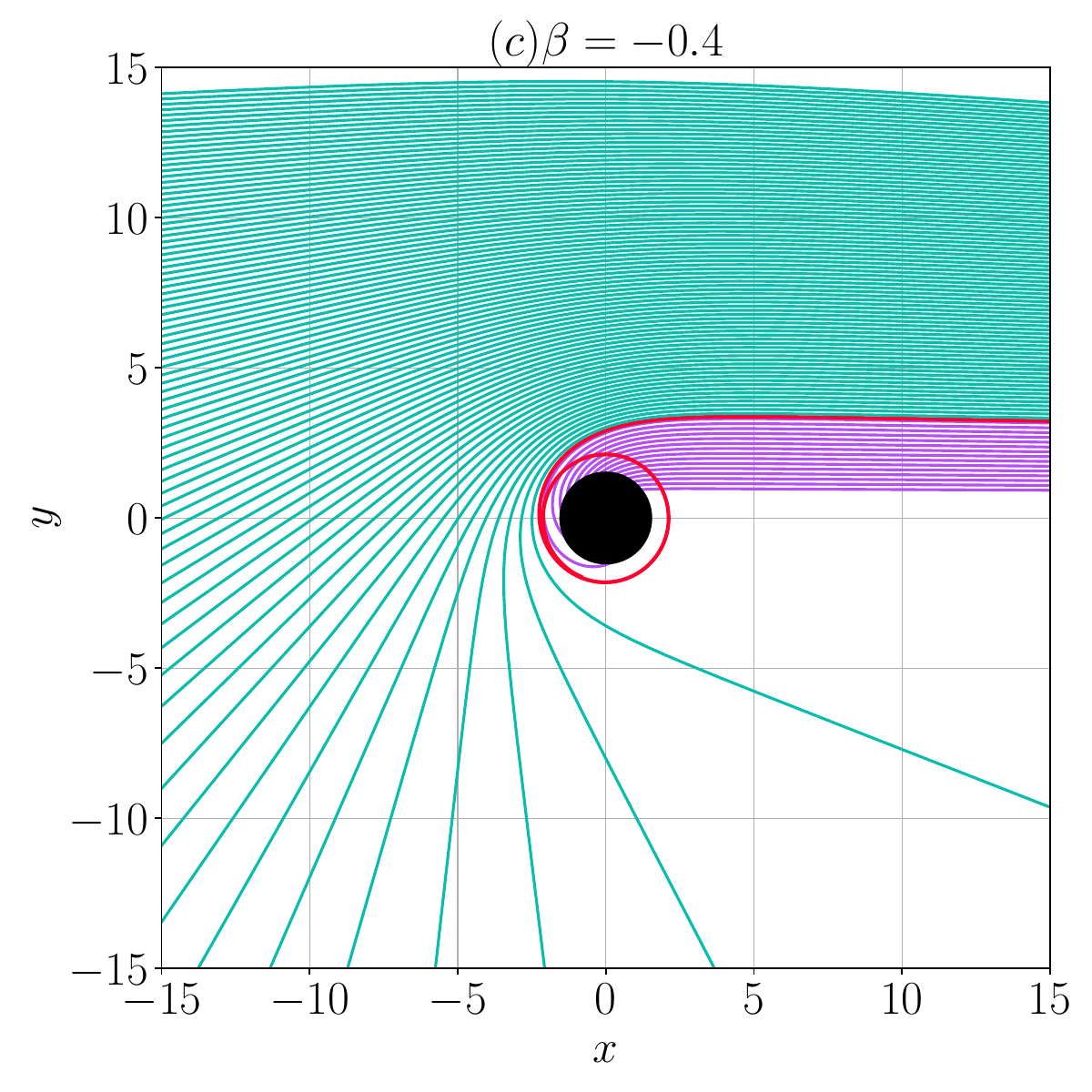}\hspace{-0.1cm}\includegraphics[scale=0.4]{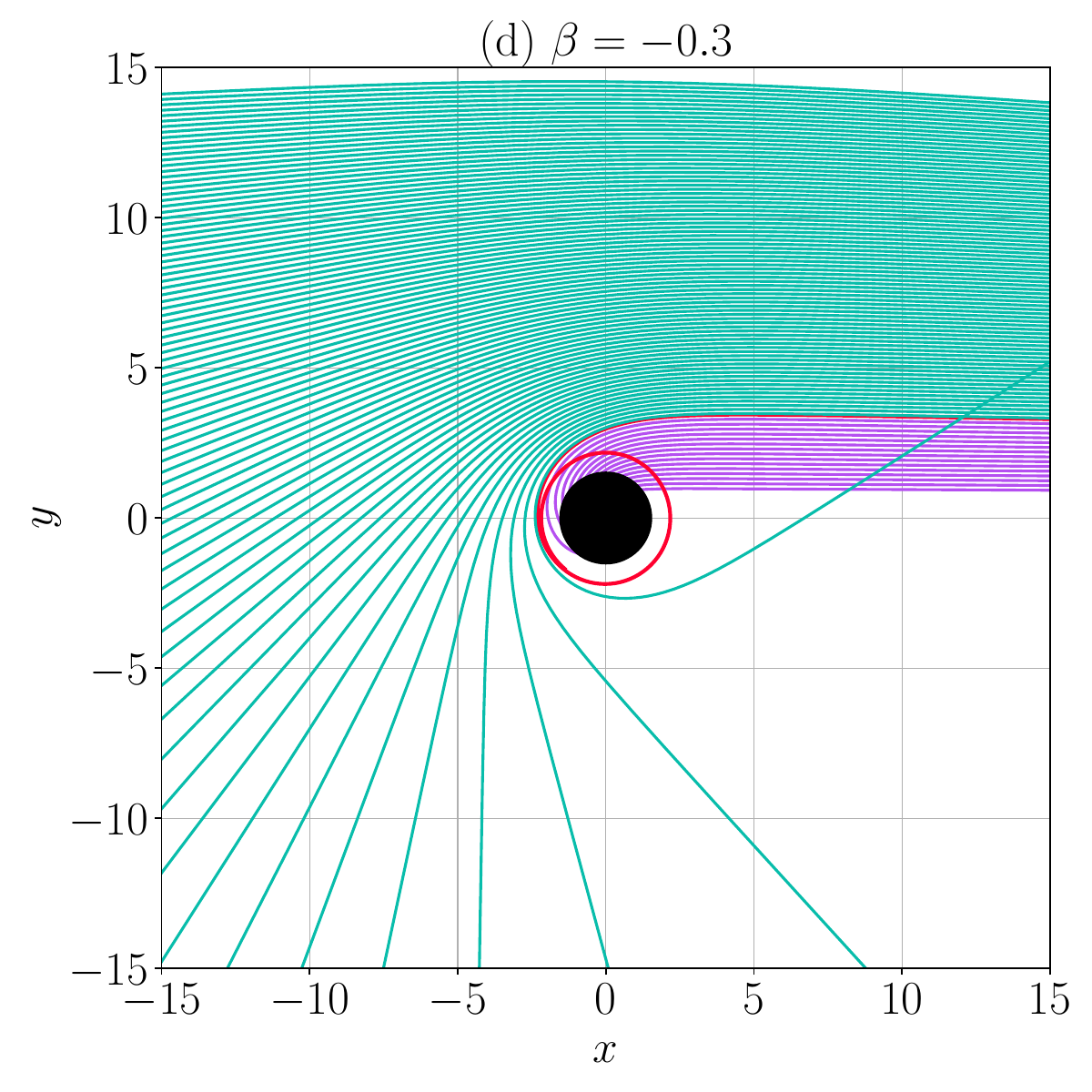}}
	\caption{Geodesics of Null rays for different values of $\beta$ with fixed $\eta = -0.5$ and $b = 0.3$.}
	\label{null_geod2}
\end{figure*}

\begin{figure*}[tbh]
	\centerline{\includegraphics[scale=0.4]{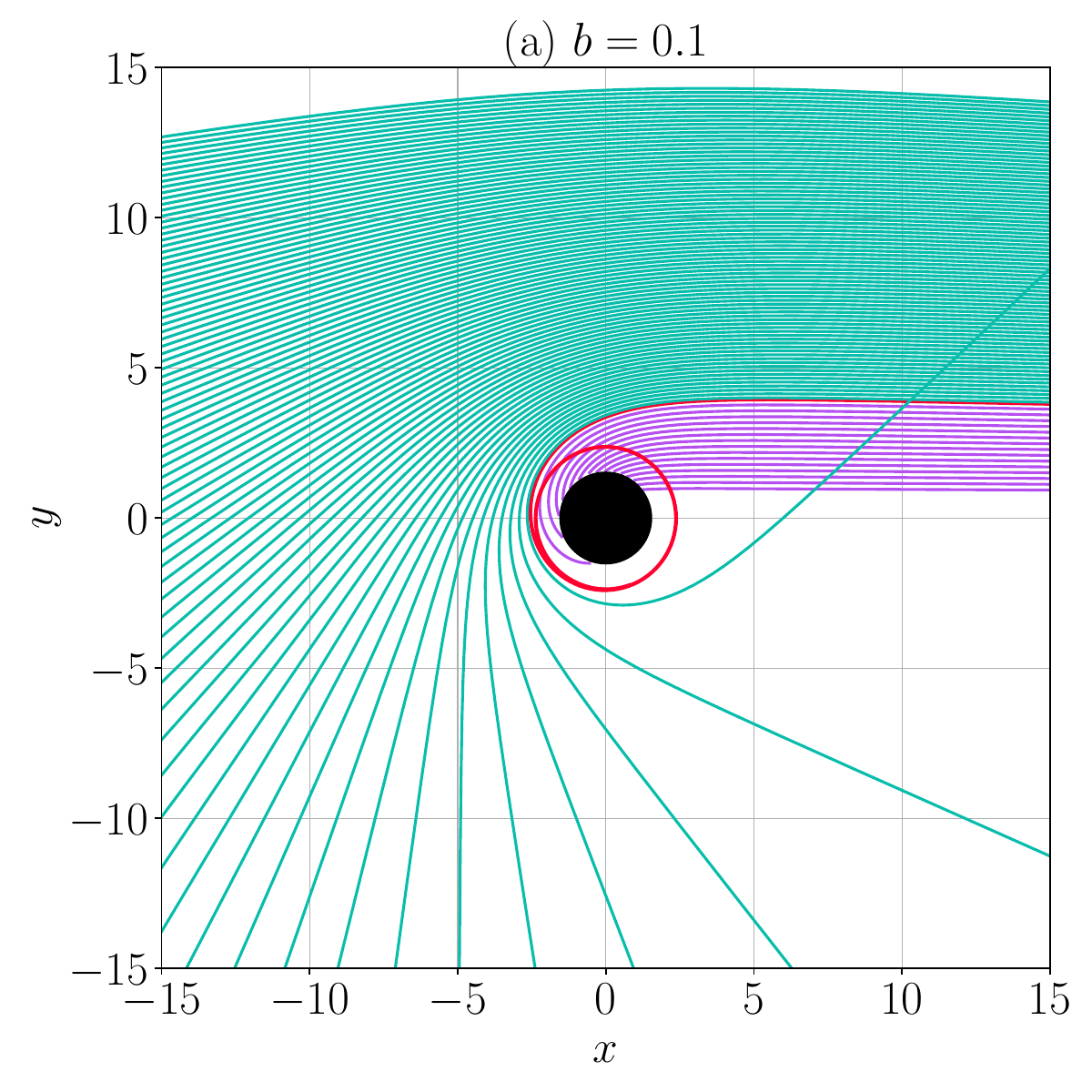}\hspace{-0.1cm}\includegraphics[scale=0.4]{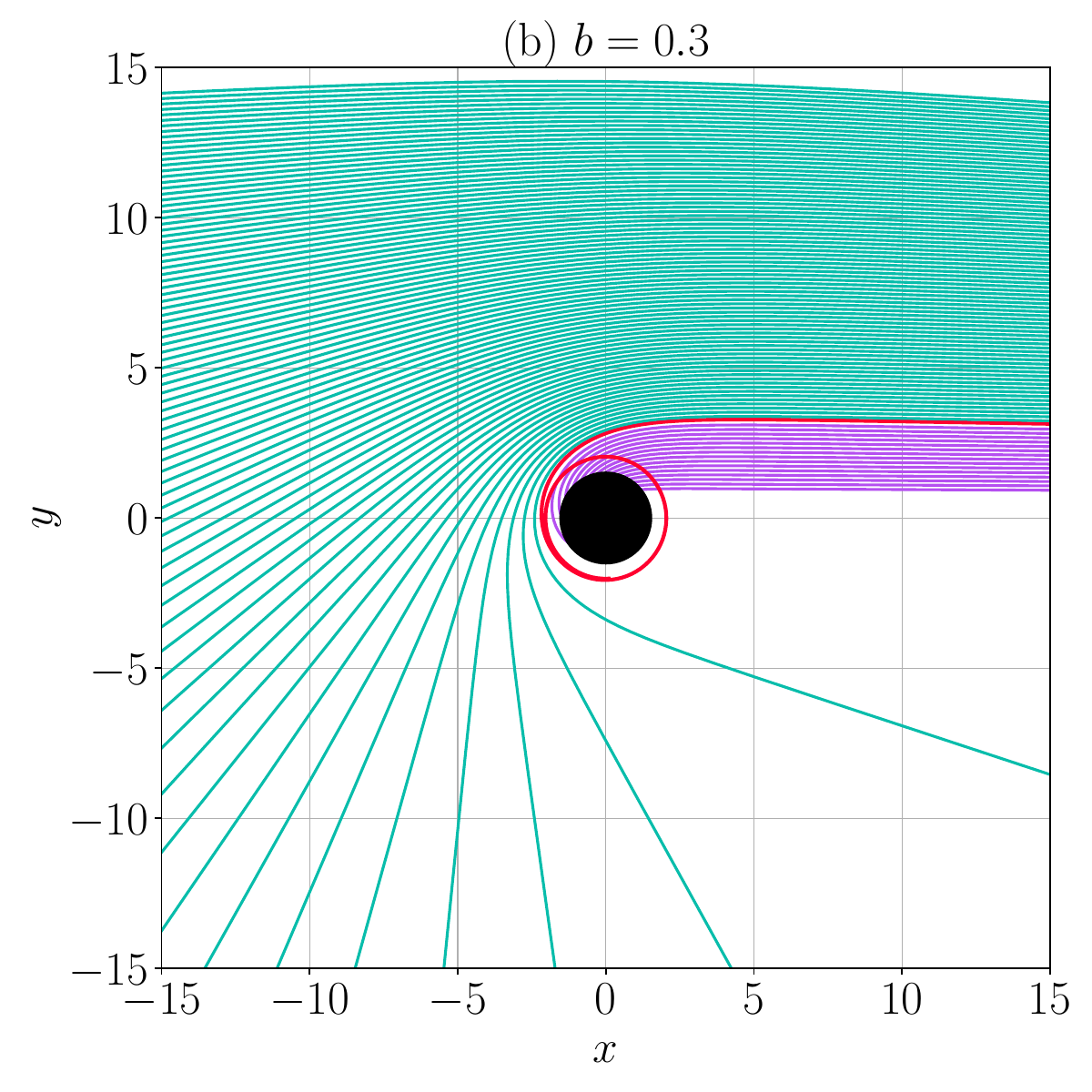}}
	\centerline{\includegraphics[scale=0.4]{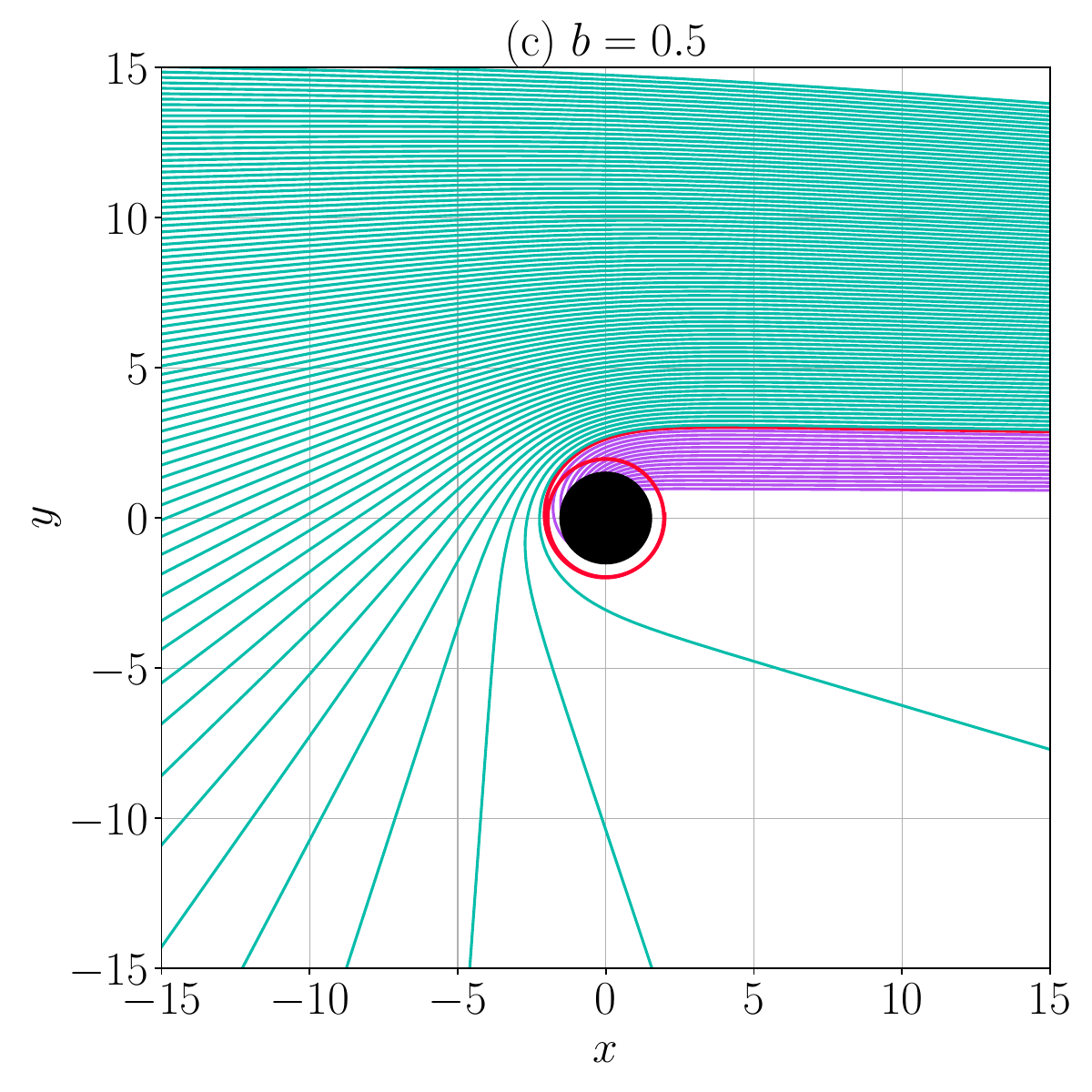}\hspace{-0.1cm}\includegraphics[scale=0.4]{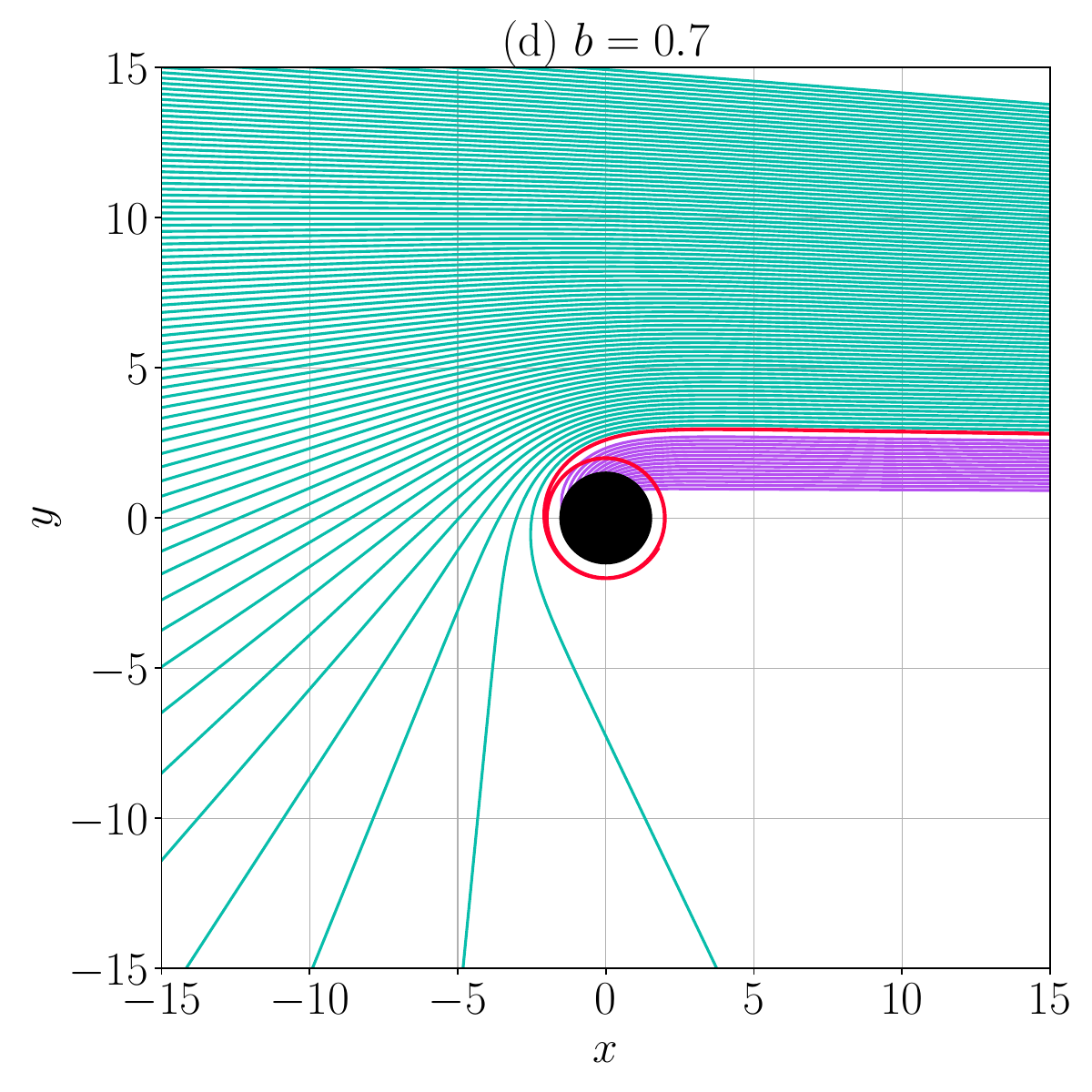}}
	\caption{Geodesics of Null rays for different values of $b$ with fixed $\beta = -0.5$ and $\eta = -0.5$.}
	\label{null_geod3}
\end{figure*}

For the null geodesics, the effective potential in our model follows from \eqref{eff2} as
\begin{equation}
V_{\rm eff}(r)
= \frac{L^2}{2r^2}\!
\biggl(1 -\frac{2M}{r} -\frac{b \ln\!\bigl(\tfrac{r}{b}\bigr)}{r}
-\frac{\beta^2}{2\eta\,r^2}\biggr),
\label{eff_pot}
\end{equation}
where \(\epsilon=0\) for null geodesics. In plotting \(V_{\rm eff}\) (Fig.~\ref{eff_pot_plot}), one may set \(E=1\) since \(E\) merely shifts the overall amplitude without altering the photon‐orbit radii.  The null geodesics themselves are obtained by numerically integrating \eqref{geod_eqs_mod}; their trajectories appear in Figs.~\ref{null_geod1},\ref{null_geod2} and \ref{null_geod3} with circular photon orbits of radii $r_p$ indicated by the red curves corresponding to the peaks of \(V_{\rm eff}\).

Panels (a), (b), and (c) of Figs. \ref{null_geod1}, \ref{null_geod2}, and \ref{null_geod3} show how the null geodesics behave using the backward ray-tracing approach when the Horndeski parameters \( \eta \), $\beta$, and dark matter parameter $b$ have different values.  The red curve in the figures represents the unstable photon orbit, which is the peak of the effective potential.  $b_{crit} = L/E$ is the critical impact parameter that correlates to this unstable photon orbit. The null rays that possess less impact parameter than $b_{crit}$ plunge into the BH while those having higher values than $b_{crit}$ deflect away from the BH (represented by green curves). The critical impact parameters and the corresponding photon radii are as tabulated in Tab. \ref{tab:grouped-params}.
\begin{table}[ht]
	\centering
	\caption{Values of \(b_{\text{crit}}\) and \(r_p\) for various combinations of \(\beta\), \(\eta\), and \(b\). Each group varies a single parameter while keeping the others fixed.}
	\label{tab:grouped-params}
	
	\begin{tabular}{
			@{} 
			c @{\hskip 1.5em} 
			c @{\hskip 1.5em} 
			c @{\hskip 1.5em} 
			c @{\hskip 1.5em} 
			c 
			@{}
		}
		\toprule
		\(\beta\) & \(\eta\) & \(b\) & \(b_{\text{crit}}\) & \(r_p\) \\
		\midrule
		\multicolumn{5}{c}{\textbf{Varying \(\eta\), fixed \(\beta = -0.5\), \(b = 0.3\)}} \\
		\midrule
		-0.5 & -0.6 & 0.3 & 3.43851 & 2.07851 \\
		-0.5 & -0.5 & 0.3 & 3.39243 & 2.04208 \\
		-0.5 & -0.4 & 0.3 & 3.32053 & 1.98483 \\
		-0.5 & -0.3 & 0.3 & 3.19175 & 1.88087 \\
		\midrule
		\multicolumn{5}{c}{\textbf{Varying \(\beta\), fixed \(\eta = -0.5\), \(b = 0.3\)}} \\
		\midrule
		-0.6 & -0.5 & 0.3 & 3.26334 & 1.93891 \\
		-0.5 & -0.5 & 0.3 & 3.39243 & 2.04208 \\
		-0.4 & -0.5 & 0.3 & 3.49029 & 2.11924 \\
		-0.3 & -0.5 & 0.3 & 3.56247 & 2.17568 \\
		\midrule
		\multicolumn{5}{c}{\textbf{Varying \(b\), fixed \(\beta = -0.5\), \(\eta = -0.5\)}} \\
		\midrule
		-0.5 & -0.5 & 0.1 & 4.09145 & 2.36405 \\
		-0.5 & -0.5 & 0.3 & 3.39243 & 2.04208 \\
		-0.5 & -0.5 & 0.5 & 3.13208 & 1.96825 \\
		-0.5 & -0.5 & 0.7 & 3.06305 & 1.99835 \\
		\bottomrule
	\end{tabular}
\end{table}

\subsection{Lyapunov Stability: Dynamical Systems Approach}
To use the Lyapunov method to analyze the stability of null circular geodesics, we construct a dynamical system and investigate its phase-space structure in the \((r,\dot{r})\) plane.   The line $(r,0)$ is the relevant phase-plane for a null circular geodesic, \(\dot r = 0\). By studying the flow in this reduced phase space, one can extract information about the critical point corresponding to the photon‐orbit radius \((r_c,0)\).  Differentiating Eq.~\eqref{eff1} and eliminating \(\dot r\) yields
\begin{equation}
\ddot{r} \;=\; -\,\frac{dV_{\rm eff}}{dr},
\label{rddot}
\end{equation}
where \(V_{\rm eff}(r)\) is given in Eq.~\eqref{eff_pot}.  Let us define
\[
x_{1} \;=\; \dot{r}, 
\qquad
x_{2} \;=\; \ddot{r} = -\,\frac{dV_{\rm eff}}{dr}.
\]
Then the dynamical system becomes:

\begin{equation}
\begin{aligned}
x_{1} &= \dot{r}, \\[6pt]
x_{2} &= -\,\frac{dV_{\rm eff}}{dr}  \\
       &= \frac{L^{2}\bigl(-\,2\beta^{2} \;+\;\eta\,r\,(b - 6M + 2r)\;-\;3b\,\eta\,r\,\ln\!\bigl(\tfrac{r}{b}\bigr)\bigr)}{2\,\eta\,r^{5}}.
\end{aligned}
\label{dyn_sys}
\end{equation}

The Jacobian matrix \(\mathcal{J}\) of the system \(\eqref{dyn_sys}\) evaluated at a generic radius \(r\) is
\begin{equation}
\mathcal{J} \;=\; 
\begin{pmatrix}
0 & 1 \\[6pt]
-\,V_{\rm eff}''(r) & 0
\end{pmatrix},
\label{jac}
\end{equation}
where \(V_{\rm eff}''(r)\equiv d^{2}V_{\rm eff}/dr^{2}\).  The characteristic equation \(\det(\mathcal{J}-\lambda\,\mathds{I})=0\) yields
\begin{equation}
\lambda^{2} \;=\; -\,V_{\rm eff}''(r).
\label{eig1}
\end{equation}
Thus:
\begin{itemize}
  \item If \(V_{\rm eff}''(r_c) > 0\), then \(\lambda^{2} < 0\), indicating a purely imaginary pair of eigenvalues and hence a \emph{stable center} at \((r_c,0)\).
  \item If \(V_{\rm eff}''(r_c) < 0\), then \(\lambda^{2} > 0\), indicating a real pair of eigenvalues of opposite sign and hence a \emph{saddle point} at \((r_c,0)\).
\end{itemize}

\begin{figure*}[tbh]
  \centering
  \includegraphics[scale=0.65]{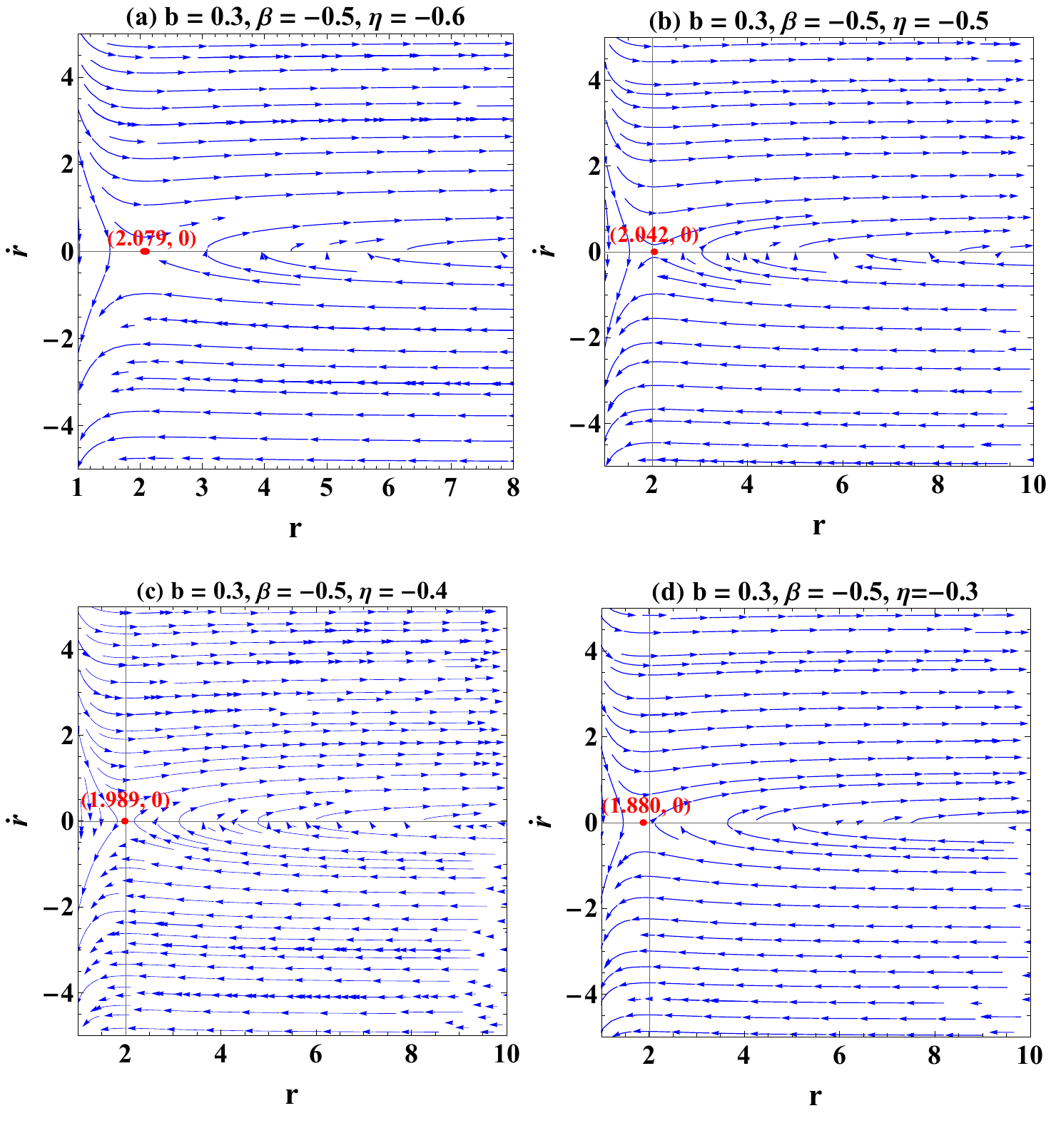}
  \caption{Phase portrait of \(r\) vs.\ \(\dot{r}\) for null geodesics, shown for different values of \(\eta\).}
  \label{phase_plot_eta}
\end{figure*}

\begin{figure*}[tbh]
  \centering
  \includegraphics[scale=0.65]{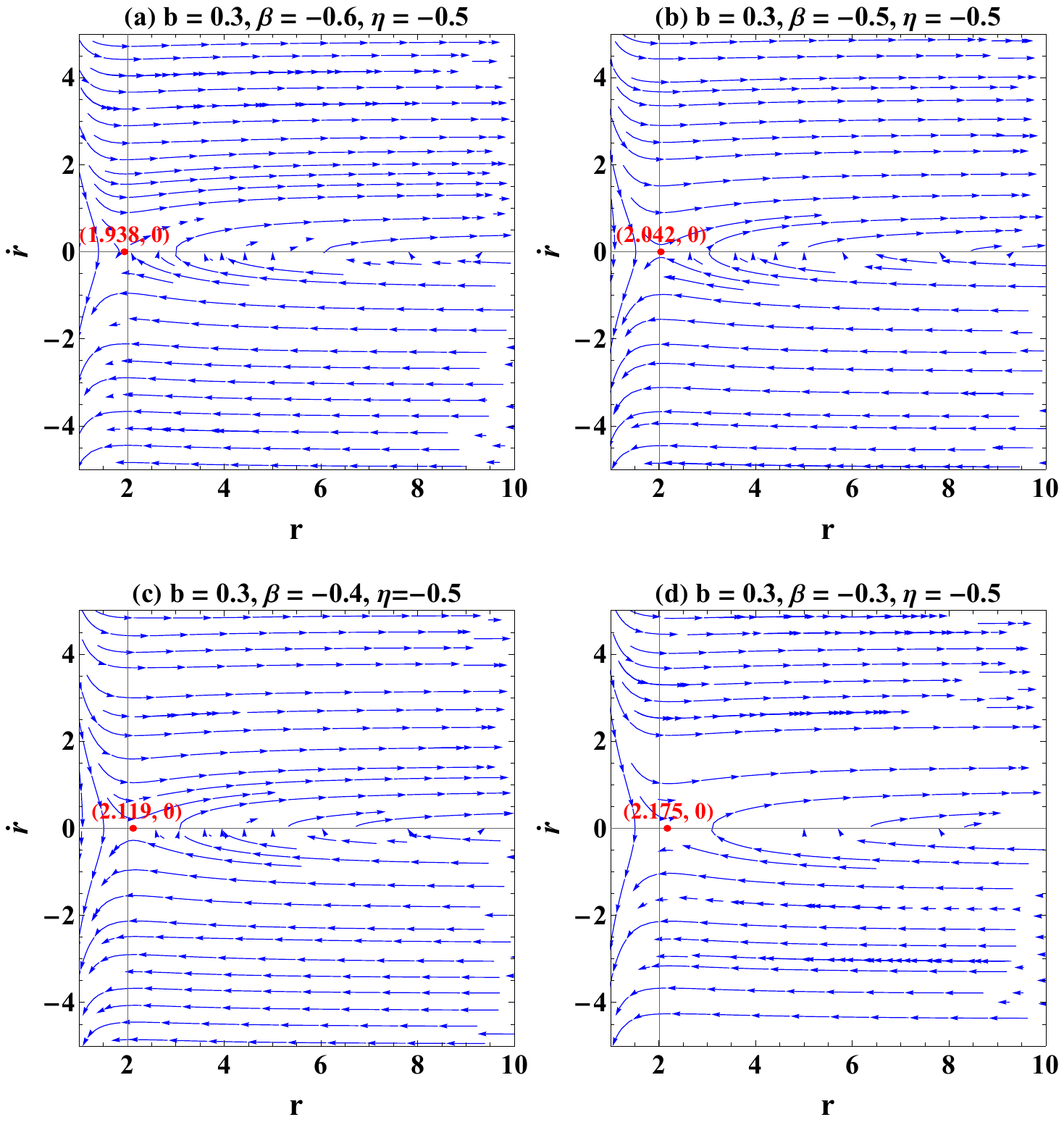}
  \caption{Phase portrait of \(r\) vs.\ \(\dot{r}\) for null geodesics, shown for different values of \(\beta\).}
  \label{phase_plot_beta}
\end{figure*}

\begin{figure*}[tbh]
  \centering
  \includegraphics[scale=0.65]{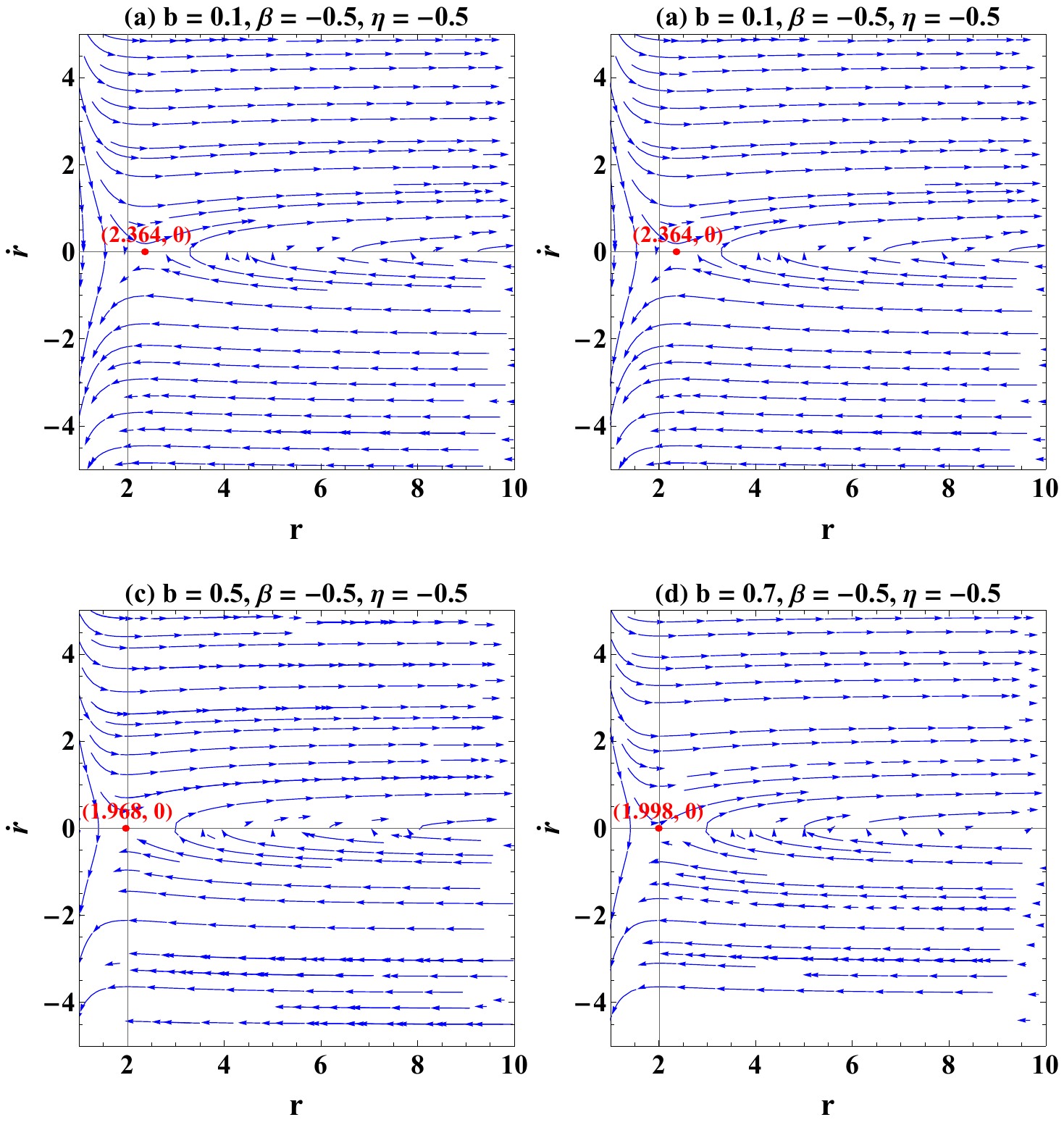}
  \caption{Phase portrait of \(r\) vs.\ \(\dot{r}\) for null geodesics, shown for different values of \(b\).}
  \label{phase_plot_b}
\end{figure*}

Figs~\ref{phase_plot_eta}–\ref{phase_plot_b} display the phase portrait \((r,\dot r)\) of the null geodesics for various values of the parameters \(\eta\), \(\beta\), and \(b\).  In each case, the photon‐sphere radius appears as the saddle critical point at the peak of the effective potential.  Any small perturbation in \(\dot r\) causes the photon to deviate from its circular orbit and fall into the BH.  This behaviour is illustrated in Fig.~\ref{phase_plot_eta}, \ref{phase_plot_beta} and \ref{phase_plot_b} for representative choices of the model parameters \(\eta\), \(\beta\), and \(b\), where the unstable photon orbits are marked as saddle points (represented by red points) on the phase plane.

\section{Shadows}
\label{sec5}
\begin{figure*}[tbh]
\centerline{\includegraphics[scale=0.55]{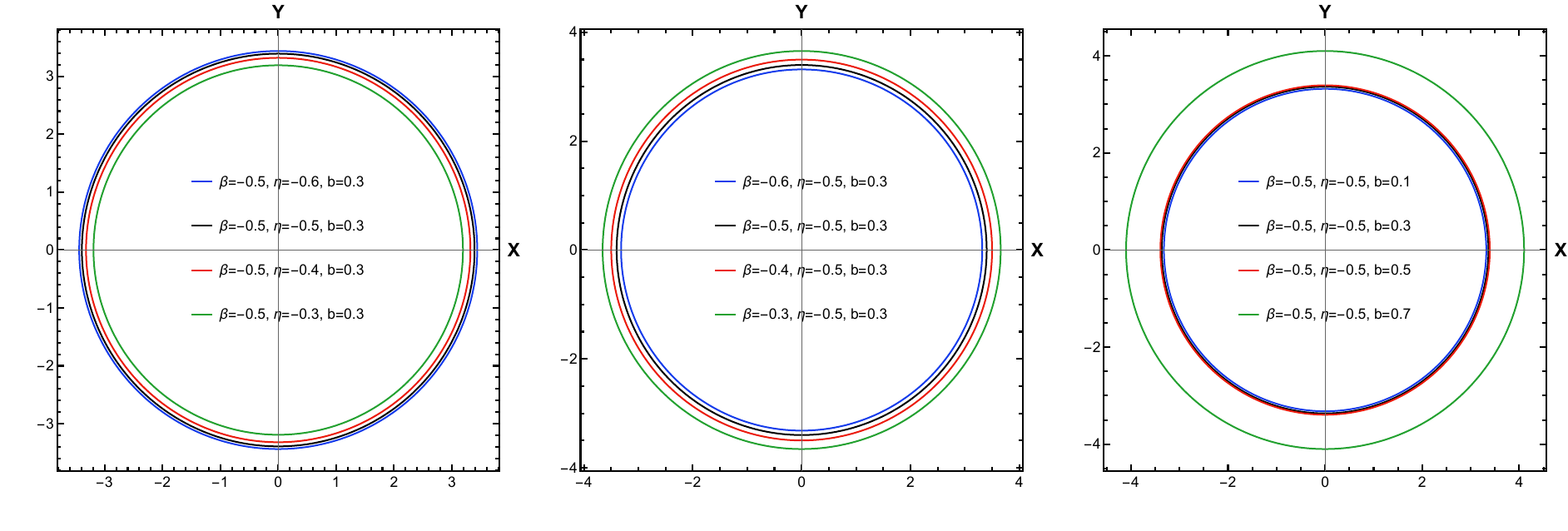}}
\caption{The plot shows the shadow images for different values of the parameters $\eta$, $\beta$, and $b$.}
\label{shadow_plot}
\end{figure*}

A BH's shadow, which contains important information about the surrounding spacetime geometry, is its most obvious observational signature.   In reality, a shadow is a black silhouette against a background of light from distant astronomical sources or an accretion disk.   The photon sphere, which is the unstable circular orbit of photons around the BH, is the main factor governing the apparent size and shape of this shadow. It is dependent on the mass, spin, and any extra fields, like PFDM in our instance.  Consequently, by comparing observed shadows to theoretical predictions, one can test general relativity under strong gravitational fields and place bounds on alternative gravity models \cite{EventHorizonTelescopeCollaboration2022May}. Now, to obtain the shadow radius,  the first step is to determine the photon orbit radius \(r_p\).  Adopting the method of Perlick and Tsupko \cite{Perlick2022Feb}, \(r_p\) satisfies
\begin{equation}
f'(r_p)\,r_p \;-\; 2f(r_p) \;=\; 0,
\label{phot_orb_rad_relation}
\end{equation}
which explicitly becomes
\begin{equation}
\frac{4 \beta ^2 + 2 \eta\,r_p\,\bigl(b + 6M - 2r_p\bigr)
- 6 b\,\eta\,r_p\,\ln\!\bigl(\frac{r_p}{b}\bigr)}
{\beta^2 - 2b\,\eta\,r_p\,\ln\!\bigl(\tfrac{r_p}{b}\bigr)
+ 2\eta\,r_p\,\bigl(2M - r_p\bigr)} = 0.
\label{phot_orb_cond_eq}
\end{equation}
Since Eq.~\eqref{phot_orb_cond_eq} cannot be solved in closed form, we obtain \(r_p\) numerically.  Once \(r_p\) is known, the shadow radius \(r_s\) follows from
\begin{equation}
r_s \;=\; \frac{r_p}{\sqrt{f(r)}\bigl|_{r_p}},
\label{shad_rad_main}
\end{equation}
as shown in Fig.~\ref{shadow_plot}.

To construct the shadow seen by a distant observer, it is convenient to introduce the celestial coordinates \(X\) and \(Y\) defined as \cite{Perlick2022Feb} :
\begin{equation}
\begin{aligned}
X &= \lim_{r_o \to \infty} \Bigl(-\,r_o^2\,\sin\theta_o \,\frac{d\phi}{dr}\Bigr),\\
Y &= \lim_{r_o \to \infty}\Bigl(r_o^2\,\frac{d\theta}{dr}\Bigr),
\end{aligned}
\label{xandy}
\end{equation}
where \(r_o\) is the observer’s radial coordinate and \(\theta_o\) is the inclination angle.  Fig ~\ref{shadow_plot} illustrates how varying \(\eta\), \(\beta\), and \(b\) alters the shadow.  In particular, increasing the $\eta$ values tends to decrease the shadow sizes, although the dependence is moderate. Similar behaviour is seen for varying $\beta$ as well. Interestingly, for varying $b$ values, the shadow size is not sensitive, but as the value is increased to 0.7, the shadow size increases drastically. This indicates, the dark matter fluid must be carefully constrained to accommodate the observed shadow data.

\subsection{Parameter constraints}
The Event Horizon Telescope’s most recent image of Sgr A$^*$ \cite{EventHorizonTelescopeCollaboration2022May} provides a unique opportunity to probe gravity in the strong-field regime.  In comparison to M87*, Sgr A$^*$’s proximity makes its mass and distance—and hence its mass-to-distance ratio—much better determined.  While M87* has a mass of order \(10^9 M_\odot\), Sgr A$^*$’s mass is of order \(10^6 M_\odot\), allowing us to test fundamental physics under even stronger curvature.  Accordingly, Sgr A$^*$’s shadow affords tighter bounds on model parameters \cite{Vagnozzi2023Jul}.

The EHT collaboration measured the fractional deviation
\begin{equation}
\delta \;\equiv\; \frac{r_s}{r_{\rm sch}} - 1
\;=\; \frac{r_s}{3\sqrt{3}\,M} - 1,
\label{del_shad}
\end{equation}
between the observed shadow radius \(r_s\) and the Schwarzschild value \(r_{\rm sch} = 3\sqrt{3}\,M\).  Keck and VLTI observations place the following constraints on \(\delta\) \cite{Do2019Jul,Abuter2020Apr,Vagnozzi2023Jul}:
\[
\delta = -0.04^{+0.09}_{-0.10} \quad (\text{Keck})
\quad\text{and}\quad
\delta = -0.08^{+0.09}_{-0.09} \quad (\text{VLTI}).
\]
Since these measurements are statistically independent, we may combine them to obtain an average
\begin{equation}
\delta \approx -0.060 \pm 0.065.
\label{av_del}
\end{equation}
Under the Gaussian approximation, the \(1\sigma\) and \(2\sigma\) intervals for \(\delta\) become
\begin{equation}
	\begin{aligned}
	&-0.125 \;\lesssim\; \delta \;\lesssim\; 0.005 
	\quad(1\sigma) \\
	& \hspace{2cm} \text{ and }  \\
	&-0.19 \;\lesssim\; \delta \;\lesssim\; 0.07 
	\quad(2\sigma).
	\end{aligned}
\label{const_aver}
\end{equation}
Rewriting Eq.~\eqref{del_shad} yields
\begin{equation}
\frac{r_s}{M} \;=\; 3\sqrt{3}\,(\delta + 1),
\label{rs_m}
\end{equation}
which, combined with Eq. \eqref{const_aver}, implies
\begin{equation}
4.55 \;\lesssim\; \frac{r_s}{M} \;\lesssim\; 5.22 \quad (1\sigma),
\quad
4.21 \;\lesssim\; \frac{r_s}{M} \;\lesssim\; 5.56 \quad (2\sigma).
\label{rs_const}
\end{equation}

\begin{figure*}[htb]
\centerline{\includegraphics[scale=0.3]{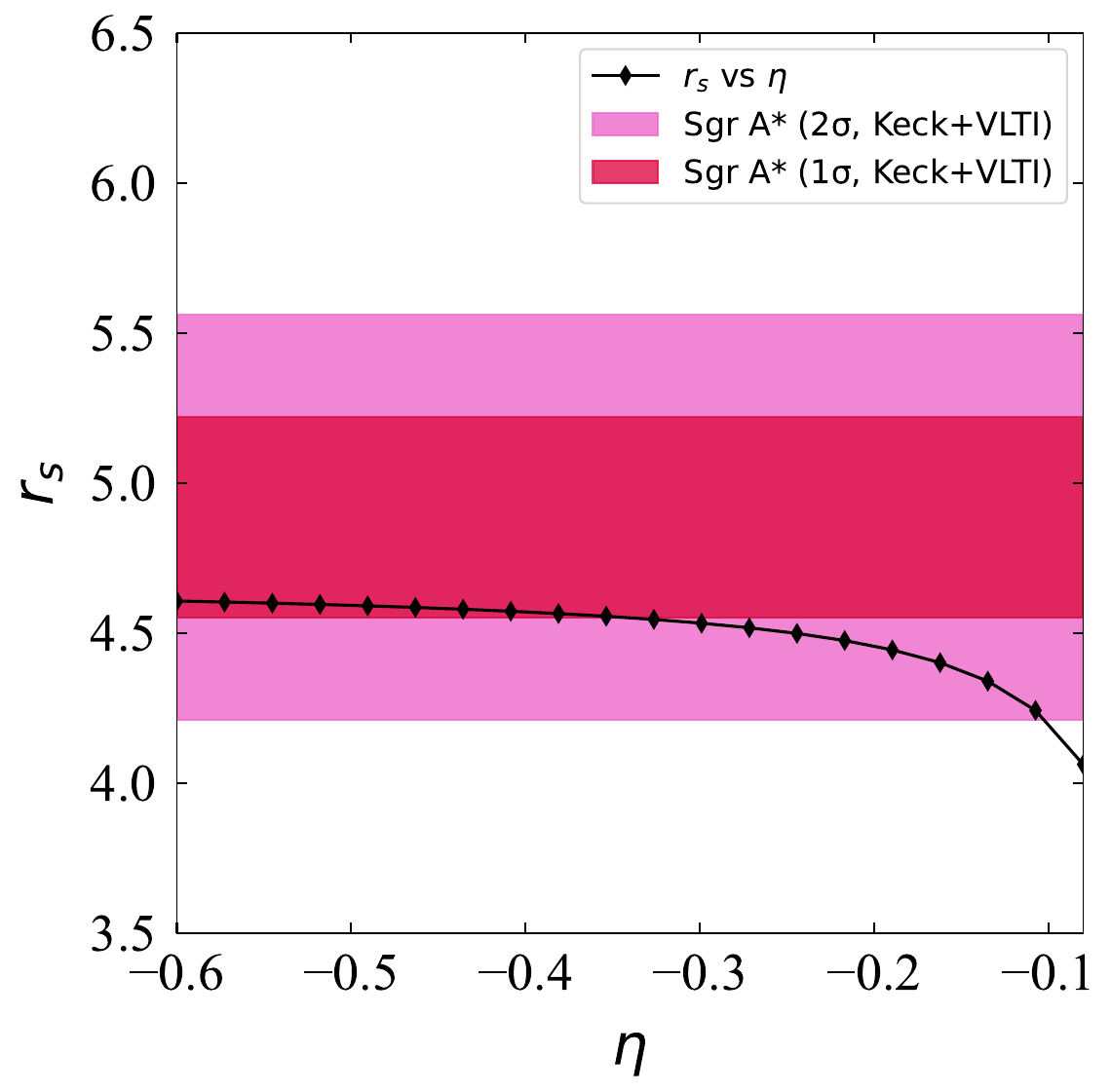}\hspace{0.2cm}\includegraphics[scale=0.3]{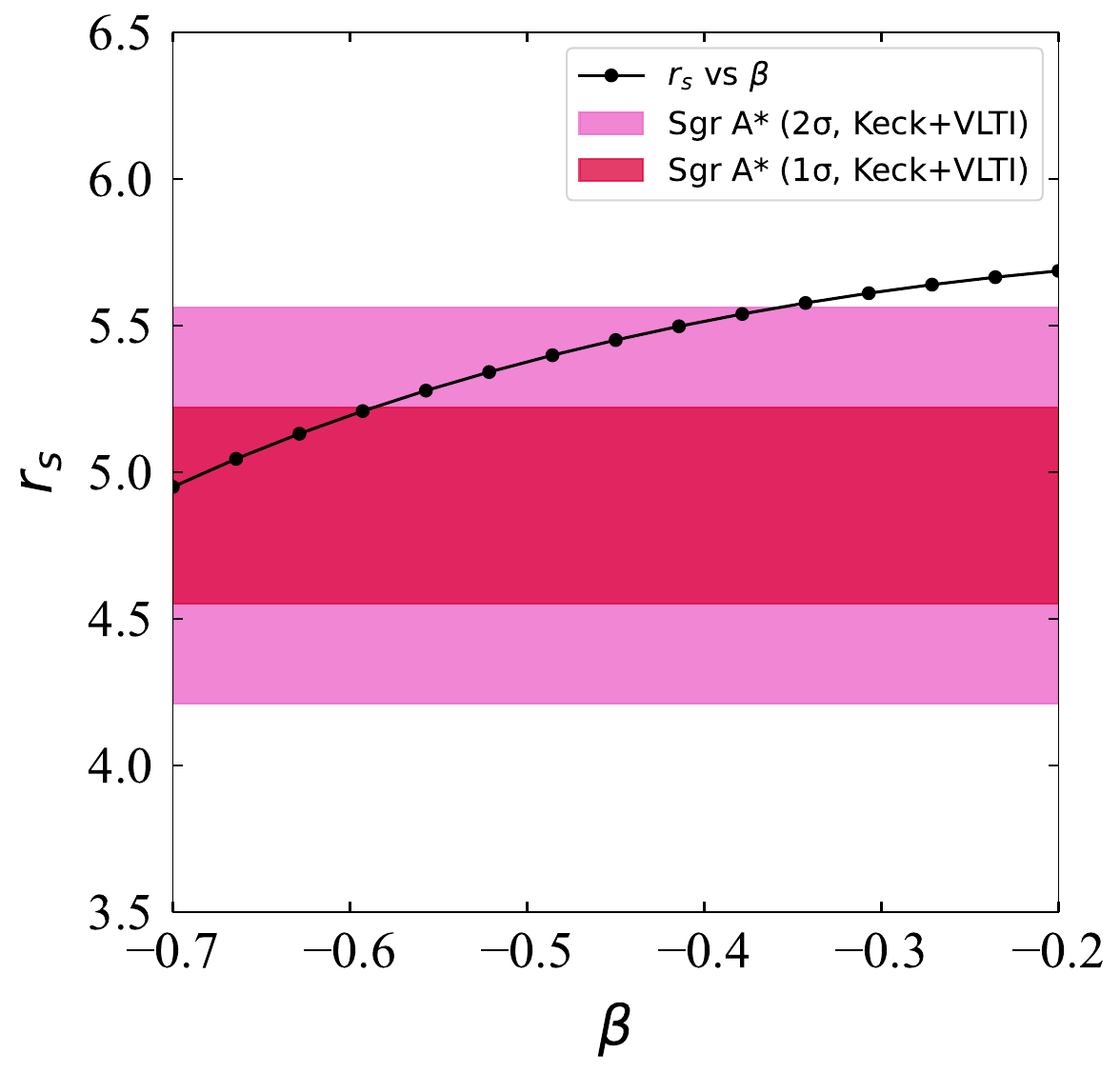}\hspace{0.2cm}\includegraphics[scale=0.3]{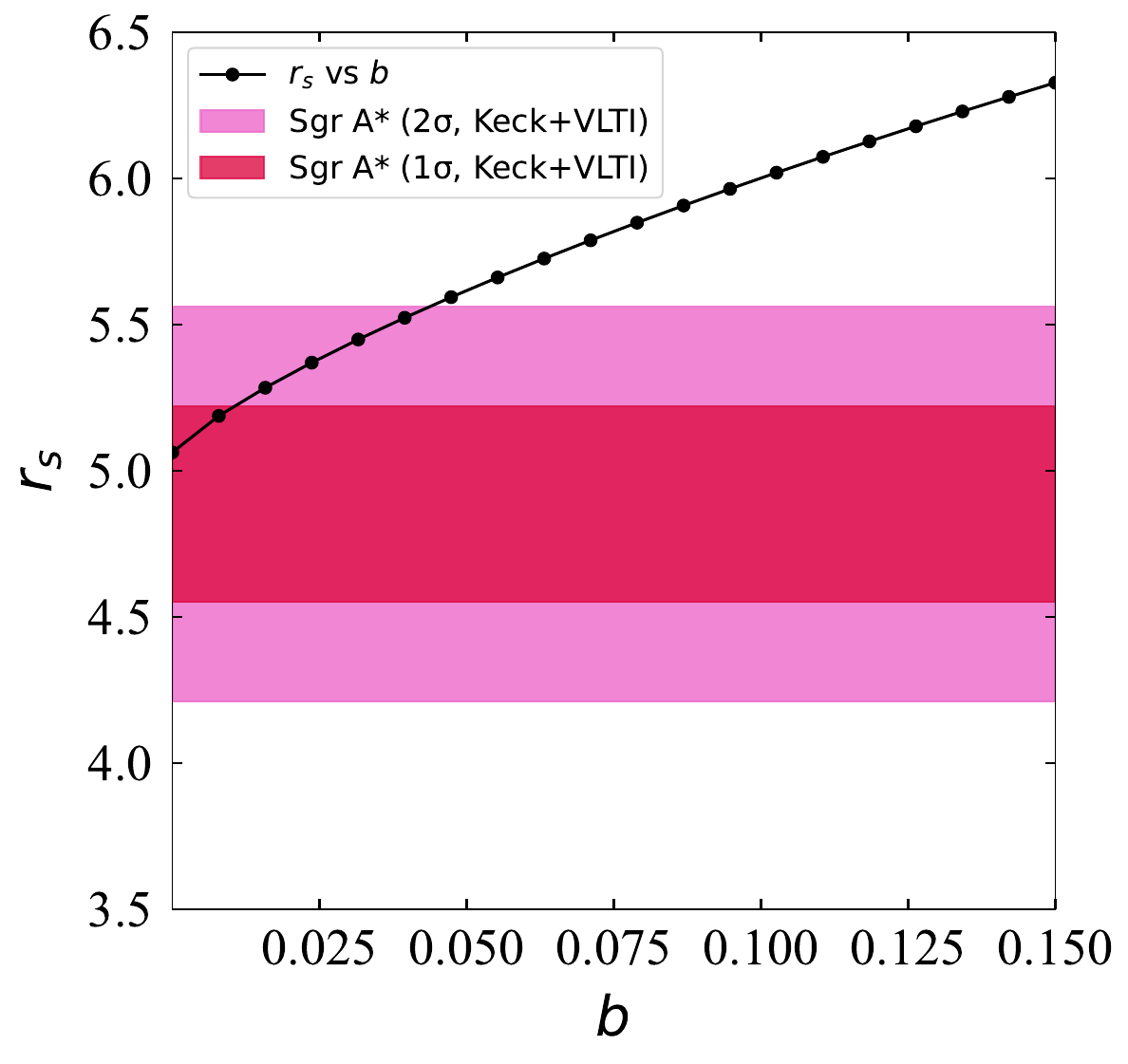}}
\centerline{\hspace{3cm} (a) \hfill \hspace{0.3cm}(b) \hfill (c) \hspace{2.2cm}}
\caption{The shadow radius is plotted against the model parameters \(\eta\), \(\beta\), and \(b\) with respect to shadow data by Keck/VLTI of Sgr A$^*$ BH. In (a), \(\beta\) and \(b\) are set to -0.5 and 0.3, in (b) \(\eta\) and \(b\) are fixed at -0.5 and 0.3, and in (c) \(\beta\) and \(\eta\) are fixed at -0.5 and -0.5, respectively.}
\label{shad_obs_plot}
\end{figure*}

Using Eqs.~\eqref{shad_rad_main} and \eqref{rs_m} together with the intervals \eqref{rs_const}, we plot \(r_s\) as a function of \(\eta\), \(\beta\), and \(b\) in Fig.~\ref{shad_obs_plot}. Based on this data, we have obtained the upper bounds on the parameters as follows: 
At the $1\sigma$ level, we obtain
\[
\beta < -0.5929,\qquad
\eta < -0.3263,\qquad
b < 0.01059.
\]
At the $2\sigma$ level, the corresponding upper bounds become
\[
\beta < -0.3429,\qquad
\eta < -0.1074,\qquad
b < 0.0436.
\]

In the PFDM extension of Horndeski gravity, these observational limits imply that any deviation from the Schwarzschild geometry due to scalar‐hair parameters $\beta$ and $\eta$ or due to a PFDM parameter $b$ must be extremely small inside the photon‐sphere region.  In particular, the requirement $b<0.01059$ at the $1\sigma$ level shows that a spherically symmetric fluid of dark‐matter density $\rho_{\rm DM}\propto b/r^{2}$ contributes negligibly to the spacetime curvature at $r\approx3M$, so that the observed shadow size is indistinguishable from the vacuum prediction.  Similarly, the bounds $\beta^2/(2\,|\eta|)\lesssim0.06$ at $1\sigma$ ensure that any Horndeski‐scalar correction to Schwarzschild case is at most a few percent near the photon sphere, again placing the metric very close to general relativity. Consequently, the current shadow constraints indicate that any PFDM component is negligible within a few gravitational radii of Sgr A*, and any Horndeski‐induced departure from Schwarzschild is sub‐percent, leaving general relativity as the effective description in that region.

\section{Quasinormal Modes}
\label{sec6}
In this section, let us discuss the properties of the QNMs of the Horndeski BH and how they are governed by the presence of the PFDM halo. First of all, let us start by studying how the BH system responds to the perturbations to the massless scalar field $\Phi$. The evolution of the scalar field is governed by the Klein-Gordon (KG) equation, which is
\begin{equation}
	\nabla_\mu \nabla^\mu \Phi = \frac{1}{\sqrt{-g}} \partial_\nu \left( g^{\mu \nu} \sqrt{-g} \partial_\mu \Phi \right) = 0,
	\label{KG eqn}
\end{equation}
Since the spacetime is spherically symmetric, we may use the method of separation of variables involving spherical harmonics $Y_{l,m} (\theta, \phi)$, with $l$ and $m$ denoting the angular and azimuthal quantum numbers, respectively. Then, the solution for $\Phi$ takes the form \cite{Fernando2012Sep,Jusufi2020Apr}
\begin{equation}
	\Phi (t, r, \theta, \phi) = \sum_{l,m} \exp (-i\omega t) Y_{l,m}(\theta, \phi) \frac{R_l(r)}{r}.
	\label{soln_KG}
\end{equation}
This solution renders the KG equation \eqref{KG eqn} to the form

\begin{equation}
	\frac{1}{r^2} \frac{d}{dr} \left[ r^2 f(r) \frac{d}{dr}\left(\frac{R(r)}{r}\right) \right] + \left( \frac{\omega^2}{f(r)} - \frac{l(l+1)}{r^2} \right) \frac{R(r)}{r} = 0.
	\label{schrod1}
\end{equation}
The above Eq. \eqref{schrod1} can be restructured into the familiar Schr\"{o}dinger like form:
\begin{equation}
	\frac{d^2 R(\ast)}{d r_{\ast}^2} + \left(\omega^2 - V(r_\ast)\right) R(r_\ast) = 0,
	\label{schrod2}
\end{equation}
where $V(r_\ast)$ is the effective potential given by:
\begin{equation}
	V(r_\ast) = f(r) \frac{l(l+1)}{r^2} + \frac{f(r)f'(r)}{r},
	\label{eff_pot1}
\end{equation}
Here, the angular quantum number $l$ can take values 0, 1, 2,\dots and $r_{\ast}$ denotes the tortoise coordinate, which is related to radial coordinate $r$ through $dr_{\ast}/dr = 1/f(r)$. It spans from $-\infty$ at the event horizon to
$+ \infty$ towards spatial infinity. It is customary to adopt the simplified notation $r_\ast \to  x, \, R(r_\ast) \to \psi(x)$ and $\omega^2 - V(r_\ast) \to Q(x)$. The potential $Q(x)$ takes constant values at $x =\pm \infty$, but not necessarily symmetric at both ends, and attains a maximum value at some point $x = x_0$. 
The Eq. \eqref{schrod2} has an exact solution when  $Q(x)$ is constant. If $Q(x)$ relies on $x$, then the equation is exactly solvable for some particular forms of $Q(x)$, and hence for $V(r_\ast)$.

Conventionally, the formulation of the WKB problem involves obtaining the solutions for incident, reflected, and transmitted amplitudes, with incident and reflected amplitudes being of equal size. In the context of BH perturbation, the incident amplitude is zero. As a result, reflected and transmitted amplitudes are found to be of the same order of magnitude. This follows that the WKB approximation is applicable given the following condition:
\begin{equation}
	\frac{d Q(x)}{dx} \Bigg|_{x = x_0} = 0.
	\label{cond_wkb}
\end{equation}
If the turning points are too close, the WKB approximation will become inapplicable. The only possible technique is to match the solutions across each turning point simultaneously. Moreover, the potential in between the two turning points is approximated to a parabolic shape \cite{Schutz1985Apr}. The matching of the solutions in the different regions is done by employing Taylor expansion of the potential about its maximum, \( x_0 \), as can be easily found in any quantum mechanics textbooks \cite{Zettili2022Sep,Griffiths2018Aug}.

The solution in the region between the turning points can be obtained by expanding \( Q(x) \) around the maximum \( x_0 \). As the solution must be continuous, it is possible to obtain the approximate solutions for the two asymptotic regions where \( Q(x) \to \omega \), a constant, at large distances. The matching conditions give what is usually known as the quasinormal mode (QNM) condition, because it leads to discrete and complex frequencies called quasinormal frequencies. The condition is given by \cite{Iyer1987Jun,Iyer1987Jun1}
\begin{equation}
	\frac{Q(x_0)}{\sqrt{2Q''(x_0)}} = - i \left(n + \frac{1}{2}\right)
	\label{condqnm}
\end{equation}
Now if one uses the condition \eqref{condqnm} in Eq. \eqref{schrod2}, it leads to
\begin{equation}
	\frac{\omega^2 - V(r_0)}{\sqrt{-2V''(r_0)}} = -i \left(n+\frac{1}{2}\right),
	\label{omeg1}
\end{equation}
or
\begin{equation}
	\omega^2 = V(r_0) - i \left( n + \frac{1}{2}\right)\sqrt{-2V''(r_0)} \equiv X - i Y.
	\label{omeg2}
\end{equation}
where $r_0$ represents the maxima of the effective potential. Here, we have also designated $X = V(r_0)$ and $Y = \left( n + \frac{1}{2}\right)\sqrt{-2V''(r_0)}$. This leads to the complex quasinormal frequencies given by
\begin{equation}
	\begin{aligned}
		&\omega = \omega_R - i \omega_I; \\& \text{ with } \quad  \omega_R = \frac{\sqrt{\sqrt{X^2+Y^2}+X}}{\sqrt{2}}, \, \\& \text{ and }  \omega_I = \frac{Y}{\sqrt{2} \sqrt{\sqrt{X^2+Y^2}+X}}
	\end{aligned}
	\label{omeg_mode}
\end{equation}

The formulation can be generalized to higher orders \cite{Iyer1987Jun,Iyer1987Jun1,Konoplya2003Jul,
	Konoplya2023Mar,Konoplya2019Jul}. 
In this work, we follow the 6$^{th}$-order WKB approximation. In the 6$^{th}$ order WKB approximation, the formula \eqref{omeg1} generalizes to 
\begin{equation}
	i \, \frac{\omega^2 - V(r_0)}{\sqrt{-2V''(r_0)}} - \sum_{j = 2}^{6} \Lambda_j = n + \frac{1}{2}
	\label{6th_wkb_omeg}
\end{equation}
The explicit correction terms $\Lambda_j$ can be found in \cite{Iyer1987Jun,Iyer1987Jun1,Konoplya2003Jul}

\begin{figure*}[htb]
	\centerline{\includegraphics[scale=0.45]{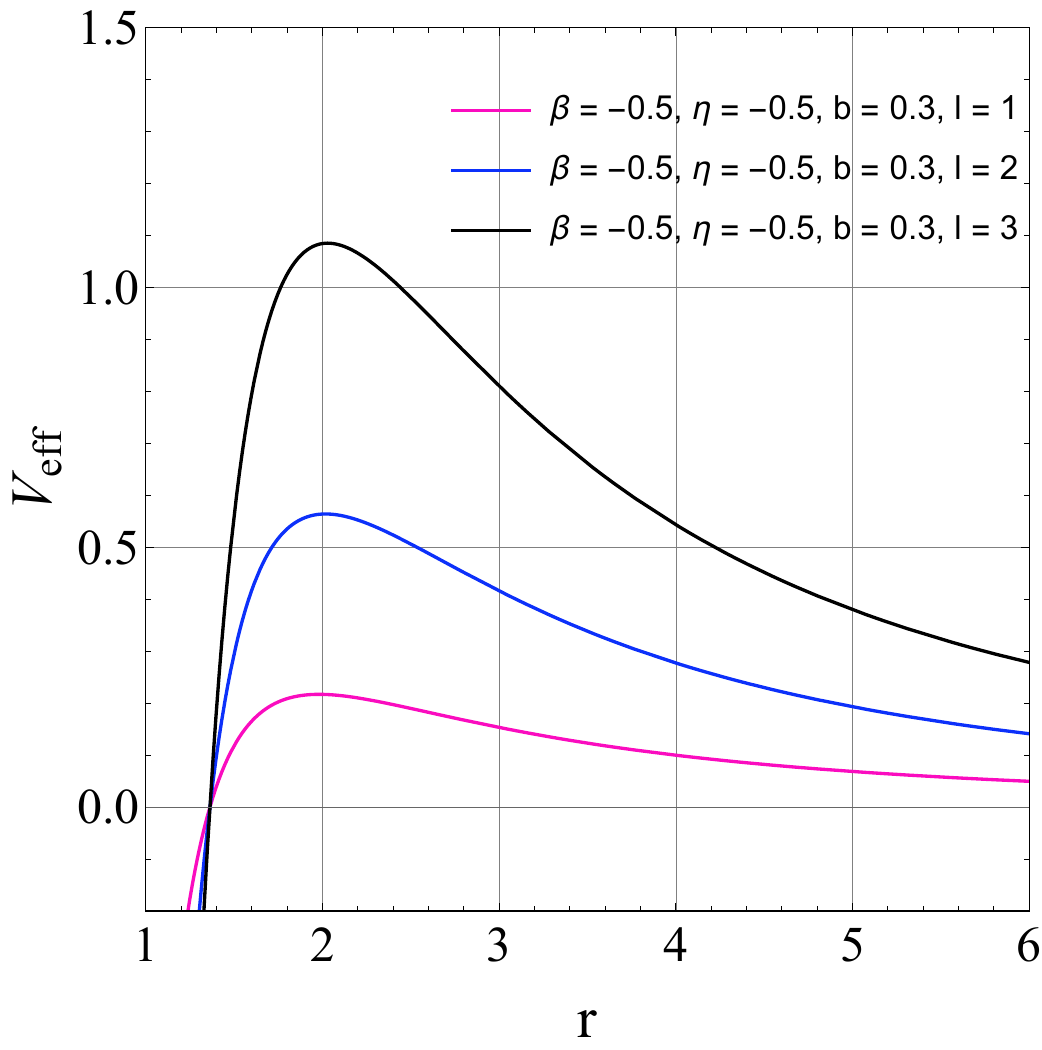}}
	\caption{The effective potential is plotted for different $l$ values for scalar modes.}
	\label{eff_pot_l}
\end{figure*}

In the Fig. \ref{eff_pot_l}, we present the radial evolution of the effective potential \(V_{\text{eff}}\) for multipole moments \(l = 1\, 2, 3\).
In the plot of \(V_{\mathrm{eff}}(r)\), as \(l\) increases from 1 to 3, the central barrier near \(r\approx3M\) becomes both higher and narrower, reflecting the familiar centrifugal term \(l(l+1)/r^2\) dominating at large \(l\). All three curves still peak close to the photon‐sphere radius. For \(l=1\), the barrier is lowest and broadest, whereas for \(l=3\) it is noticeably taller and steeper, causing higher‐\(l\) modes to probe a more concentrated potential well. At large \(r\), all potentials fall off like \(1/r^2\), and near the horizon they approach zero, since \(V_{\mathrm{eff}}(r_h)=0\) where the lapse function vanishes.


\begin{table*}[ht]
  \centering
  \caption{Quasinormal mode frequencies \(\omega_{l}\) for different values of \(\eta\).}
  \label{tab:qnm-omegas}

  \begin{tabular}{
    @{} 
    c @{\hskip 1.5em} 
    c @{\hskip 1.5em} 
    c @{\hskip 1.5em} 
    c 
    @{}
  }
    \toprule
    \(\eta\) 
      & \(\omega_{l=1}\) 
      & \(\omega_{l=2}\) 
      & \(\omega_{l=3}\) \\
    \midrule
    \(-0.6\) 
      & \(0.420264 - 0.153745\,i\) 
      & \(0.717446 - 0.153698\,i\) 
      & \(1.01098  - 0.153778\,i\) \\
    \(-0.5\) 
      & \(0.427621 - 0.154632\,i\) 
      & \(0.728150 - 0.154473\,i\) 
      & \(1.02539  - 0.154523\,i\) \\
    \(-0.4\) 
      & \(0.439653 - 0.155894\,i\) 
      & \(0.745530 - 0.155562\,i\) 
      & \(1.04874  - 0.155568\,i\) \\
    \(-0.3\) 
      & \(0.463059 - 0.157637\,i\) 
      & \(0.778904 - 0.157012\,i\) 
      & \(1.09338  - 0.156947\,i\) \\
    \bottomrule
  \end{tabular}

\end{table*}

\begin{table*}[ht]
  \centering
  \caption{Quasinormal mode frequencies \(\omega_{l}\) for different values of \(\beta\).}
  \label{tab:qnm-omegas-beta-variant}

  \begin{tabular}{
    @{} 
    c @{\hskip 1.5em} 
    c @{\hskip 1.5em} 
    c @{\hskip 1.5em} 
    c 
    @{}
  }
    \toprule
    \(\beta\) 
      & \(\omega_{l=1}\) 
      & \(\omega_{l=2}\) 
      & \(\omega_{l=3}\) \\
    \midrule
    \(-0.6\) 
      & \(0.449737 - 0.156765\,i\) 
      & \(0.759979 - 0.156298\,i\) 
      & \(1.06809  - 0.156270\,i\) \\
    \(-0.5\) 
      & \(0.427621 - 0.154632\,i\) 
      & \(0.728150 - 0.154473\,i\) 
      & \(1.02539  - 0.154523\,i\) \\
    \(-0.4\) 
      & \(0.412299 - 0.152692\,i\) 
      & \(0.705795 - 0.152769\,i\) 
      & \(0.995271 - 0.152882\,i\) \\
    \(-0.3\) 
      & \(0.401702 - 0.151144\,i\) 
      & \(0.690189 - 0.151392\,i\) 
      & \(0.974183 - 0.151550\,i\) \\
    \bottomrule
  \end{tabular}

\end{table*}

\begin{table*}[ht]
  \centering
  \caption{Quasinormal mode frequencies \(\omega_{l}\) for different values of \(b\).}
  \label{tab:qnm-omegas-b}

  \begin{tabular}{@{} 
    c @{\hskip 1.5em} 
    c @{\hskip 1.5em} 
    c @{\hskip 1.5em} 
    c 
    @{}}
    \toprule
    \(b\) 
      & \(\omega_{l=1}\) 
      & \(\omega_{l=2}\) 
      & \(\omega_{l=3}\) \\
    \midrule
    \(0.1\) 
      & \(0.352207 - 0.120290\,i\) 
      & \(0.601594 - 0.120252\,i\) 
      & \(0.847853 - 0.120297\,i\) \\
    \(0.3\) 
      & \(0.427621 - 0.154632\,i\) 
      & \(0.728150 - 0.154473\,i\) 
      & \(1.025390 - 0.154523\,i\) \\
    \(0.5\) 
      & \(0.459940 - 0.178425\,i\) 
      & \(0.787826 - 0.178562\,i\) 
      & \(1.111140 - 0.178747\,i\) \\
    \(0.7\) 
      & \(0.464224 - 0.192399\,i\) 
      & \(0.803168 - 0.193092\,i\) 
      & \(1.135700 - 0.193499\,i\) \\
    \bottomrule
  \end{tabular}
\end{table*}

\begin{table}[htb]
  \centering
  \caption{Rounded relative errors \(\Delta_{6,l}\) (mantissa to two decimal places) for QNM frequencies at various \(\eta_{1}\), \(\beta_{1}\), and \(b\).}
  \label{tab:qnm-errors-rounded}

  \begin{tabular}{
    @{} 
    c @{\hskip 1.5em} 
    c @{\hskip 1.5em} 
    c @{\hskip 1.5em} 
    c 
    @{}
  }
    \toprule
    \(\eta_{1}\) 
      & \(\Delta_{6,\;l=1}\) 
      & \(\Delta_{6,\;l=2}\) 
      & \(\Delta_{6,\;l=3}\) \\
    \midrule
    \(-0.6\) & \(3.72\times10^{-4}\) & \(2.61\times10^{-5}\) & \(4.30\times10^{-6}\) \\
    \(-0.5\) & \(3.69\times10^{-4}\) & \(2.64\times10^{-5}\) & \(4.45\times10^{-6}\) \\
    \(-0.4\) & \(3.68\times10^{-4}\) & \(2.72\times10^{-5}\) & \(4.83\times10^{-6}\) \\
    \(-0.3\) & \(2.83\times10^{-4}\) & \(3.03\times10^{-5}\) & \(5.53\times10^{-6}\) \\
    \bottomrule
  \end{tabular}

  \vspace{1ex}

  \begin{tabular}{
    @{} 
    c @{\hskip 1.5em} 
    c @{\hskip 1.5em} 
    c @{\hskip 1.5em} 
    c 
    @{}
  }
    \toprule
    \(\beta_{1}\) 
      & \(\Delta_{6,\;l=1}\) 
      & \(\Delta_{6,\;l=2}\) 
      & \(\Delta_{6,\;l=3}\) \\
    \midrule
    \(-0.6\) & \(3.33\times10^{-4}\) & \(3.04\times10^{-5}\) & \(5.35\times10^{-6}\) \\
    \(-0.5\) & \(3.69\times10^{-4}\) & \(2.64\times10^{-5}\) & \(4.45\times10^{-6}\) \\
    \(-0.4\) & \(3.95\times10^{-4}\) & \(2.51\times10^{-5}\) & \(4.04\times10^{-6}\) \\
    \(-0.3\) & \(4.14\times10^{-4}\) & \(2.43\times10^{-5}\) & \(3.75\times10^{-6}\) \\
    \bottomrule
  \end{tabular}

  \vspace{1ex}

  \begin{tabular}{
    @{} 
    c @{\hskip 1.5em} 
    c @{\hskip 1.5em} 
    c @{\hskip 1.5em} 
    c 
    @{}
  }
    \toprule
    \(b\) 
      & \(\Delta_{6,\;l=1}\) 
      & \(\Delta_{6,\;l=2}\) 
      & \(\Delta_{6,\;l=3}\) \\
    \midrule
    \(0.1\) & \(2.01\times10^{-4}\) & \(1.47\times10^{-5}\) & \(2.43\times10^{-6}\) \\
    \(0.3\) & \(3.69\times10^{-4}\) & \(2.64\times10^{-5}\) & \(4.45\times10^{-6}\) \\
    \(0.5\) & \(6.27\times10^{-4}\) & \(4.37\times10^{-5}\) & \(7.18\times10^{-6}\) \\
    \(0.7\) & \(1.02\times10^{-3}\) & \(6.28\times10^{-5}\) & \(9.86\times10^{-6}\) \\
    \bottomrule
  \end{tabular}

\end{table}
\begin{figure*}[htb]
	\centerline{\includegraphics[scale=0.6]{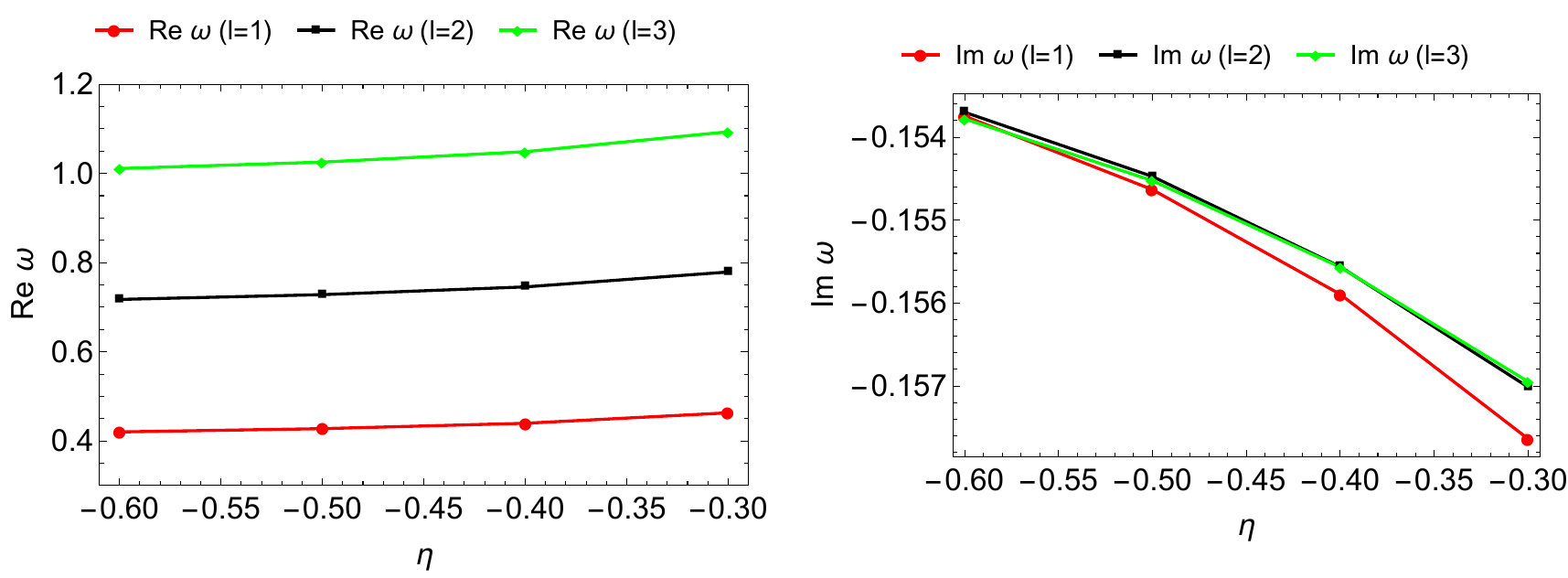}}
	\caption{Plot of real and imaginary components for varying $\eta$ with $\beta$ and $b$ fixed at $-0.5$ and $0.3$ respectively.}
	\label{qnm_plot_eta}
\end{figure*}

\begin{figure*}[htb]
	\centerline{\includegraphics[scale=0.6]{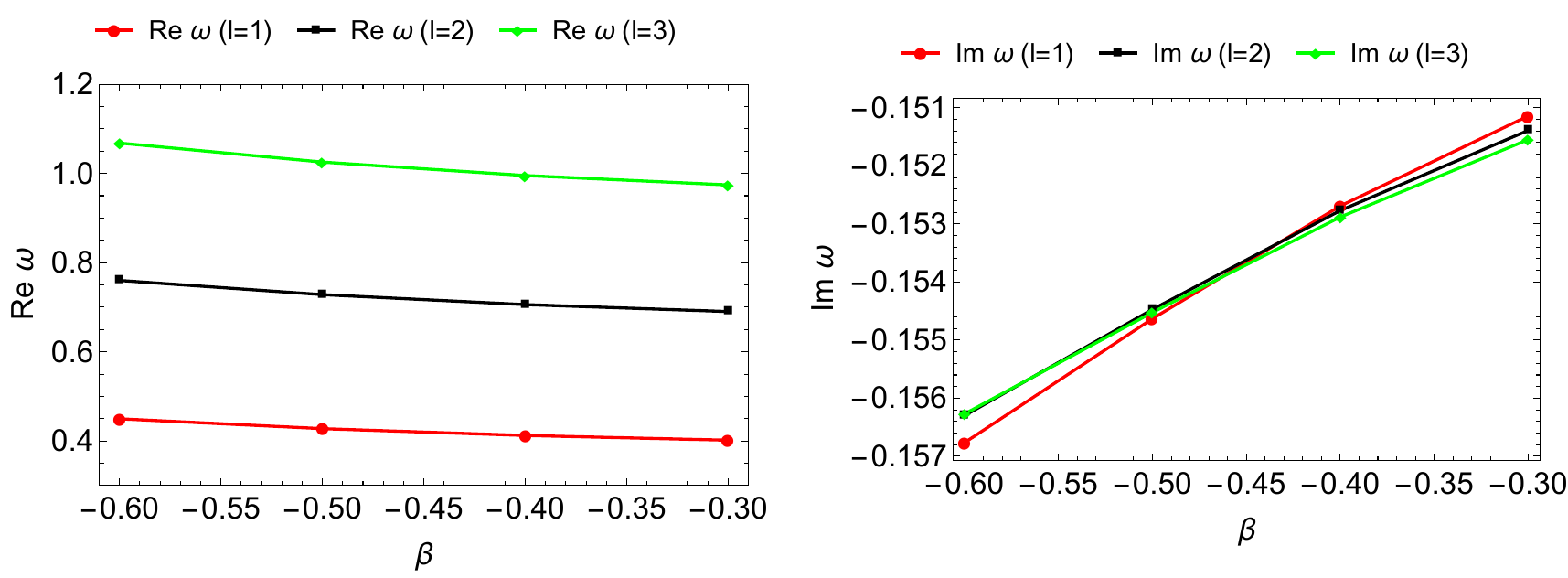}}
	\caption{Plot of real and imaginary components for varying $\beta$ with $\eta$ and $b$ fixed at $-0.5$ and $0.3$ respectively.}
	\label{qnm_plot_beta}
\end{figure*}


\begin{figure*}[htb]
	\centerline{\includegraphics[scale=0.6]{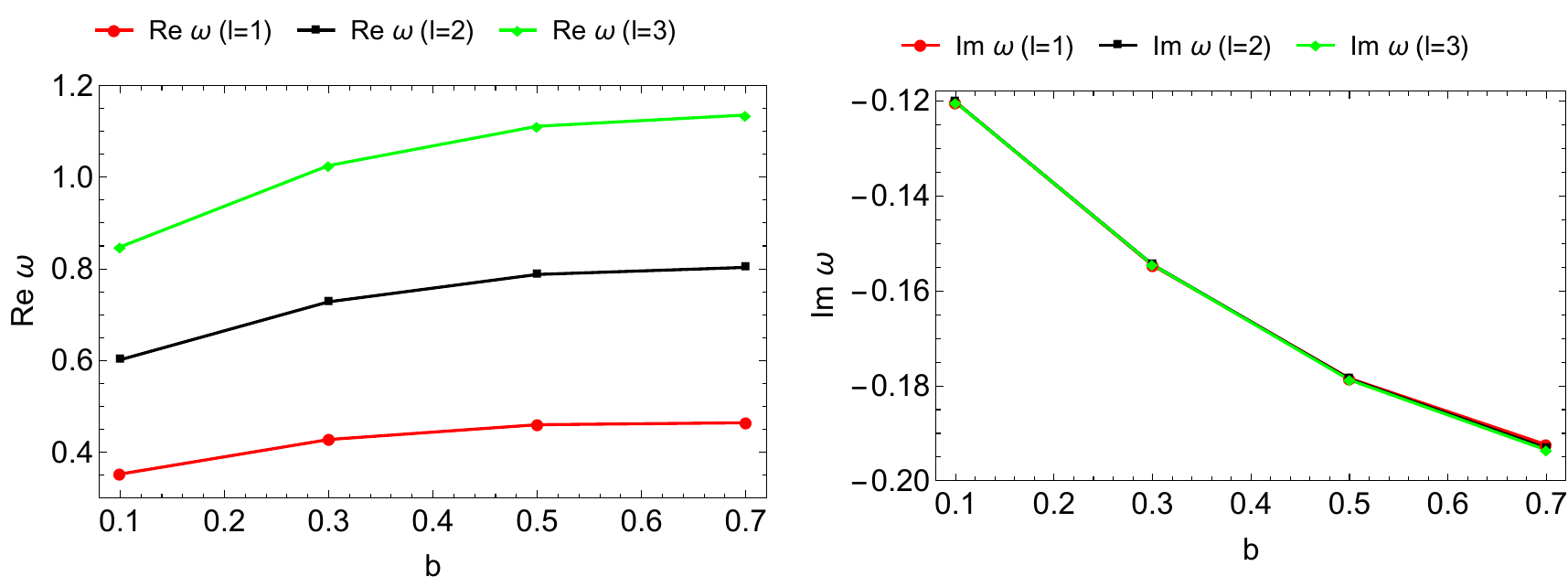}}
	\caption{Plot of real and imaginary components for varying $b$ with $\eta$ and $\beta$ fixed at $-0.5$ and $-0.5$ respectively.}
	\label{qnm_plot_b}
\end{figure*}

In Table~\ref{tab:qnm-omegas}, we see that increasing the Horndeski parameter $\eta_{1}$ from $-0.6$ to $-0.3$ causes both parts of the fundamental quasinormal‐mode frequency for $\ell=1$ to grow: the real part rises from $0.420264$ to $0.463059$, and the imaginary part increases in magnitude from $0.153745$ to $0.157637$.  Similar trends appear for $\ell=2$ ($|\Re|$: $0.717446\to0.778904$, $|\Im|\!:0.153698\to0.157012$) and $\ell=3$ ($|\Re|$: $1.010980\to1.093380$, $|\Im|\!:0.153778\to0.156947$).  In other words, making $\eta$ less negative raises both the oscillation frequency and the damping rate of the modes.
By contrast, Table~\ref{tab:qnm-omegas-beta-variant} shows that as $\beta$ increases from $-0.6$ to $-0.3$, the real part of $\omega_{\ell=1}$ decreases from $0.449737$ to $0.401702$ and the magnitude of its imaginary part drops from $0.156765$ to $0.151144$, with $\ell=2$ and $\ell=3$ following the same pattern.  Here, a more negative $\beta$ strengthens the tensor coupling in $G_{4}=1+\beta\sqrt{-X}$, so both $\Re(\omega)$ and $|\Im(\omega)|$ grow.  Conversely, reducing $|\beta|$ lowers both parts of the frequency.
Table~\ref{tab:qnm-omegas-b} illustrates the effect of the PFDM $b$.  As $b$ goes from $0.1$ to $0.7$, the real part $\Re(\omega_{\ell=1})$ increases from $0.352207$ to $0.464224$, while $|\Im(\omega_{\ell=1})|$ rises from $0.120290$ to $0.192399$.  Similar increases occur for $\ell=2$ ($0.601594\to0.803168$, $|\Im|\!:0.120252\to0.193092$) and $\ell=3$ ($0.847853 \to 1.135700$, $|\Im|\!:0.120297\to0.193499$).  A larger $b$ tends to boost both the oscillation frequency and the damping rate. These results are portrayed in Figs.~\ref{qnm_plot_eta}, \ref{qnm_plot_beta} and \ref{qnm_plot_b}.
Finally, Table~\ref{tab:qnm-errors-rounded} shows that the rounded relative errors $\Delta_{6,\ell}$ at sixth‐order WKB are very small.  For changes in $\eta$, $\Delta_{6,\ell=1}$ stays around $10^{-4}$ (e.g.\ $3.72\times10^{-4}\to2.83\times10^{-4}$ as $\eta$ goes from $-0.6$ to $-0.3$) and $\Delta_{6,\ell=3}$ stays around $10^{-6}$ ($4.30\times10^{-6}\to5.53\times10^{-6}$).  When $\beta$ varies, $\Delta_{6,\ell=1}$ changes from $3.33\times10^{-4}$ to $4.14\times10^{-4}$ and $\Delta_{6,\ell=3}$ from $5.35\times10^{-6}$ down to $3.75\times10^{-6}$.  For $b$ between $0.1$ and $0.7$, $\Delta_{6,\ell=1}$ grows from $2.01\times10^{-4}$ to $1.02\times10^{-3}$, while $\Delta_{6,\ell=3}$ remains below $10^{-5}$.  These small errors confirm that the sixth‐order WKB method converges well, even when external DM or Horndeski corrections are present.

Taken together, the data confirm that (i) a less negative $\eta$, raises both $\Re(\omega)$ and $\lvert\Im(\omega)\rvert$  (ii) a less negative $\beta$ weakens the Horndeski tensor coupling, lowering frequencies and damping rates; and (iii) increasing $b$ produce higher oscillation frequencies and faster damping. Moreover, the sixth‐order WKB approximation remains reliable (relative errors $\lesssim10^{-3}$ for $\ell=1$, $\lesssim10^{-5}$ for $\ell=3$) across these parameter ranges.


\section{Conclusion}
\label{sec7}
In summary, we have presented a detailed analysis of a quartic square‐root Horndeski BH embedded in a PFDM surrounding. In the first part of our paper, a thermodynamic analysis reveals a joint role of the Horndeski scalar coupling parameters and PFDM parameters on the BH properties. The local stability transition point is shifted towards a higher horizon radius as the PFDM parameter \(b\) effect becomes stronger. Thus, it is seen that in this model, local stability is shown by smaller BHs rather than the larger ones. Another crucial result of this work is that the strictly positive Helmholtz free energy indicates that this system evades any Hawking-Page transition, suggesting that PFDM-surrounded BHs remain globally unstable yet locally quasi-stable over an extended radius range. Second, we studied the null-geodesics by solving the geodesic equations by implementing the backward ray-tracing method and Lyapunov phase space stability analysis to find out how the presence of scalar hair and PFDM subtly reshape the photon orbits.  We demonstrate that increasing \(\lvert\beta\rvert\) or \(b\) consistently enlarges the critical impact parameter and photon sphere radius, while making \(\eta\) less negative contracts these orbits. The unstable saddle nature of these circular trajectories underpins a mechanism by which the Horndeski terms and dark matter jointly destabilize the null trajectories.
Third, by confronting our theoretical shadow predictions with EHT measurements of Sgr A$^*$, we derived the tight bounds on these non-minimal Horndeski couplings and PFDM parameter. These bounds suggest that any PFDM-induced effects or Horndeski scalar hair within the photon sphere region must be very small.  To the best of our knowledge, this represents a novel quantitative constraint on PFDM parameter in a Horndeski context using BH shadow data.
Finally, we calculated the quasinormal modes of the BH via the sixth-order WKB approximation method. These led us to the following results: both the oscillation frequencies and damping rates grow with PFDM density and with less negative \(\eta\), but diminish as \(\lvert\beta\rvert\) decreases.  The exceptionally small relative errors confirm the reliability of the WKB predictions. These results shows a promising possibility for gravitational wave astronomy to probe scalar-tensor and dark-matter effects concurrently. Nevertheless, based on the shadow constraints, we expect that the Horndeski modified gravity contribution should be quite minimal.

Future work could extend our analysis to include rotating versions of Horndeski–PFDM BHs and other dark-matter halo profiles, thereby refining the predictions for shadows and ringdown signals. Incorporating plasma effects or accretion-disk physics would further bridge theory and observations.

\bibliography{bibliography.bib}

\end{document}